\documentclass[11pt,twoside]{article} 
\usepackage[utf8]{inputenc}
\usepackage[a4paper, margin=1in]{geometry}
\usepackage{authblk} 
\usepackage{fancyhdr}
\usepackage{lipsum} 

\usepackage[english]{babel}

\usepackage{authblk}

\usepackage[colorlinks=true, allcolors=blue]{hyperref}

\usepackage{cite}
\usepackage{amsmath,amssymb,amsfonts}
\usepackage{algorithmic}
\usepackage{graphicx}
\usepackage{textcomp}
\usepackage{pdfpages}
\usepackage{xcolor}
\usepackage{subcaption}

\usepackage{float}
\usepackage{comment}
\usepackage{xspace}

\PassOptionsToPackage{hyphens}{url}
\usepackage{hyperref}
\usepackage[colorinlistoftodos]{todonotes}
\usepackage{verbatim}

\title{Toward Space-Based Public Key Systems: Enabling Secure Space Communications through In-Orbit Trust Services}

\author[1]{Rehana Yasmin\thanks{e-mail: rehana.yasmin@kaust.edu.sa}}
\author[2]{Paulo Esteves-Verissimo\thanks{e-mail: pjverissimo@ciencias.ulisboa.pt}}
\author[1]{Ali Shoker\thanks{e-mail: ali.shoker@kaust.edu.sa}}

\affil[1]{King Abdullah University of Science and Technology, 
Saudi Arabia}
\affil[2]{LASIGE, Faculdade de Ciências, Universidade de Lisboa, Portugal}

\date{}

\begin{document}
\maketitle


\begin{abstract}
 
The “New Space” era has led to a rapid increase in satellites operated by independent entities in near-Earth orbit. This shift enables richer space services but also requires secure, near-real-time coordination, making efficient authentication of space assets critical for next-generation missions. Traditional ground-dependent Public Key Infrastructure (PKI) suffers from latency and operational bottlenecks that limit scalability and availability in dynamic space environments. This paper proposes architectural designs for space-based PKI that shift certificate management and validation from ground infrastructure into space, reducing reliance on ground stations while enabling interoperability and cross-entity collaboration. Two deployment schemes are introduced: a space-ground integrated PKI with in-orbit validation authorities, and a fully autonomous space-based PKI with in-space issuance and validation. We analyze deployment trade-offs in scalability, availability, security, cost, and operational complexity in multi-operator environments. A baseline latency analysis is provided to illustrate performance implications of in-orbit trust management.
 
\end{abstract}


\section{Introduction}
\label{intro}

Future next-generation space missions are moving from isolated communication systems toward interconnected networks of space-based entities with heterogeneous nodes \cite{CCSDS350.6-G-1}. This vision of a fully integrated spacecraft ecosystem, where satellites from different operators communicate directly and exchange information, aligns with NASA’s concept for next-generation Earth science missions \cite{intercomm2001}. Realizing such a space ecosystem requires secure communication and robust trust management across multiple operators.

Low Earth Orbit (LEO) satellites are gaining increasing importance as their proximity to Earth (160-2,000 km) enables applications such as Earth observation, broadband Internet, and disaster recovery, tasks that are more difficult for higher-altitude systems like geostationary (GEO) satellites at 35,780 km \cite{NASAOrbits-2009}. However, LEO satellites move at high speeds, which limits ground station visibility and continuous communication due to orbital dynamics. Medium Earth Orbit (MEO), which lies between LEO and GEO, is also increasingly used in modern satellite systems and often operates alongside LEO constellations. Line-of-sight limitations, together with environmental factors such as rain attenuation, further degrade reliable data exchange with ground stations in LEO and MEO.

To address this, satellites increasingly rely on Inter-Satellite Communication (ISC), which enables direct communication between satellites across different constellations and orbital layers. ISC reduces dependence on ground stations and helps avoid delays caused by congestion, limited line-of-sight, and environmental factors \cite{Yates2023,ESA}. It also enables real-time decision making, which is critical for applications such as disaster monitoring, collision avoidance, and space traffic management. In addition, cross-operator satellite networks can provide global coverage, while mitigating traffic congestion and enabling more efficient use of limited orbital resources. ISC also facilitates emerging applications such as space-based data centers  \cite{ESA-2024,ASCEND-2024}, federated satellite systems \cite{GOLKAR2015230,VONMAURICH2018}, and software-defined missions \cite{Loft-2024}. These capabilities inherently require secure authentication, trust establishment, and protected communication to ensure safe and reliable operations among heterogeneous nodes.

As space networks become more open and interconnected, security concerns grow significant. Data, commands, and telemetry control signals can be intercepted or manipulated by adversaries. An attacker may inject false commands, take control of a satellite, misuse its functions, or even force it out of orbit, potentially jeopardizing the entire mission, or other stakeholders' operations. Strong \textit{authentication}, data \textit{integrity} of commands and telemetry, and \textit{confidentiality} of sensitive data are therefore essential to prevent unauthorized access, command corruption, and information leakage, which can lead to loss of control at the ground station, spacecraft failure, or exposure of sensitive mission data \cite{CCSDS351.0-M-1}.

Cryptographic mechanisms, including symmetric and asymmetric key techniques, are recommended by the Consultative Committee for Space Data Systems (CCSDS) for securing space systems \cite{CCSDS, CCSDS352.0-B-2}. However, as satellite networks scale to hundreds or thousands of heterogeneous nodes, key management becomes a major challenge. Symmetric key approaches do not scale efficiently due to extensive key distribution requirements, while public key systems offer better scalability but introduce challenges in certificate management and validation \cite{CCSDS350.6-G-1}. Traditional PKI assumes reliable, low-latency, always-connected terrestrial infrastructure, which does not hold in space environments characterized by intermittent connectivity, long delays, and limited onboard resources; for example, a satellite may be unable to contact a certificate authority during a communication window to validate a certificate. As space systems evolve, quantum-resistant cryptography is expected to become increasingly relevant, requiring PKI designs that support migration to post-quantum algorithms without changing their underlying architecture.

In current space deployments, satellites rely on certificates issued by different ground-based PKI systems \cite{CCSDS350.6-G-1}, requiring validation across multiple certificate authorities. Preloading certificates before launch is one option, but it is constrained by onboard storage and lacks flexibility for dynamic environments. Alternatively, certificates can be transmitted to ground for validation; however, this introduces higher latency due to intermittent connectivity and reliance on relay systems, while increasing dependence on ground infrastructure. Certificate-based authentication remains challenging in highly dynamic and distributed space networks due to constrained onboard resources, multi-CA coordination, intermittent connectivity, and requirements for autonomous verification. These limitations indicate that traditional ground-centric PKI approaches are insufficient in space environments.

To address these challenges, this paper makes the following contributions:

\begin{itemize}
    \item \textit{\textbf{Space-based PKI architecture with in-orbit trust services:}} We propose a \textit{novel system-level space-based PKI architecture} that treats \textit{certificate issuance and validation as in-orbit services} rather than ground-only functions. This reduces reliance on ground stations and enables interoperability and cross-agency collaboration in space environments.
    
    \item \textit{\textbf{Two deployment schemes for in-space PKI operations:}} We present (i) an \textit{incremental scheme} where only certificate validation is performed in space while issuance remains on the ground, and (ii) a \textit{fully in-space scheme} where both issuance and validation are performed in orbit. These schemes define alternative deployment models for space-based PKI.
      
    \item \textit{\textbf{System-level feasibility and deployment analysis:}} We \textit{examine feasibility} across orbital configurations, system design, scalability, and security, and analyze the impact of moving trust services from ground to space in terms of \textit{deployment choices} and operational complexity. We further \textit{discuss operational and business considerations} in multi-operator and multi-agency space environments.
    
    \item \textit{\textbf{Reference workflow and communication latency evaluation:}} We provide a reference workflow for PKI operations in space environments and evaluate communication latency to establish a baseline for future implementations, supporting the practical grounding of the proposed architecture.
\end{itemize}

The rest of the paper is structured as follows: Section~\ref{PKI_Space} discusses current PKI practices, the need for in-orbit PKI, and implementation challenges. Section~\ref{background} presents the preliminaries. Section~\ref{PKI} details the proposed PKI deployment approaches, including design considerations and latency evaluation with comparisons to existing methods. Section~\ref{related} reviews related work. Section~\ref{future-work} outlines open challenges and future directions. Finally, Section~\ref{conclusion} concludes the paper.


\section{PKI for Space Security}
\label{PKI_Space} 

Traditional certificate verification in ground-based networks relies on CA communication, which is not directly applicable in space. Thus, space networks rely on alternative methods.

\subsection{Current Practices in Certificate Deployment and Validation}
\label{AuthPath}

In current deployments, a trusted CA issues certificates to satellites and ground stations. Before launch, each satellite is preloaded with its own certificate and, in some cases, a small set of trusted root or intermediate certificates. During validation, a satellite may attempt to verify a certificate using its stored trust anchors; however, due to limited onboard storage, it cannot maintain a complete set of trusted certificates and revocation lists, which restricts its ability to perform full validation independently. Alternatively, the satellite forwards the certificate to a ground station for validation, including chain and revocation checks, and receives the result in return. Since, LEO and MEO satellites are not in direct line of sight of ground always, certificate validation may be accomplished in one of the following ways: 
\begin{enumerate}

    \item \textit{Delayed Validation} where satellites must wait until they next pass an approved ground station before performing validation.
    
    \item \textit{Near Real-Time Validation via Data Relay Satellites} where LEO and MEO satellites send certificates to a GEO relay satellite. Geo satellite, in a circular orbit, remains stationary over the equator, matching Earth’s rotation and maintaining continuous visibility to ground. Thus, a GEO satellite acts as a relay for LEO/MEO satellites which forwards certificate to a ground station for validation and relays the results back along the same path. A practical example of a relay system is NASA's Tracking and Data Relay Satellites system \cite{TDRS}, which provides a constant communication link for LEO and MEO satellites.
\end{enumerate}

Both approaches introduce latency, with relay-based validation further incurring transmission and processing overhead along the relay path as discussed in Sections \ref{validtion_Delayed} and \ref{validtion_Relay}. Reducing latency and reliance on ground stations, can strengthen security and support interoperability across satellite networks. However, current research is limited, and existing systems are often tied to single agencies or missions, hindering cross-agency collaborations. 


\subsection{Need for Space-Based Public Key Systems}

As inter-satellite communication becomes more widespread, ensuring the security of these links through mutual authentication is increasingly critical. The growing complexity, autonomy, and scale of satellite networks, along with emerging use cases, drive the need for space-based public key systems. Relying solely on ground-based PKI is inefficient and impractical; space-based PKI systems can manage and verify certificates directly in orbit, enabling secure and efficient inter-satellite communication. They can extend trust across organizations and support joint missions, satellite constellations, and emerging applications, as well as global challenges such as space traffic management and collaborative missions.

\subsubsection{Reduced Dependency on Ground Stations}

Space-based PKI systems reduce reliance on ground-based systems, which is \textit{a key motivation for their adoption}. In-space certificate verification helps avoid delays caused by \textit{ground station unavailability} and \textit{communication disruptions}, while enhancing the \textit{autonomy of satellite networks}.

\begin{itemize}

    \item \textit{Overcoming Ground Station Unavailability:} Ground stations may experience congestion during high-traffic periods when multiple satellites require communication simultaneously, a challenge particularly evident for stations with limited capacity or those supporting large constellations or multiple operators. They may also become unavailable due to their geographically restricted coverage, as satellites can only communicate when within the station's line of sight. Additionally, physical and cyber threats, including natural disasters, signal jamming, or cyberattacks targeting ground station operations, can result in unavailability. Space-based PKI systems overcome ground station unavailability by enabling secure communication and authentication directly between satellites, reducing reliance on terrestrial infrastructure, ensuring continuity of secure operations in scenarios where ground communication is delayed or unavailable.
        
    \item \textit{Resilience to Communication Disruptions:} Rain attenuation, or rain fade, significantly impacts communication between satellites and ground stations by absorbing and scattering electromagnetic signals, which results in signal degradation or loss \cite{Yates2023}. It also introduces noise into the communication system, especially for higher frequency bands commonly used in satellite communications \cite{ESA}. Space-based PKI systems provide an additional layer of resilience against communication disruptions between satellites and ground stations, ensuring continuous secure operations even when ground links are compromised or unavailable.

    \item \textit{Increased Autonomy of Satellite Networks:} Reducing dependency on ground stations elevates satellite network autonomy by allowing satellites to process data onboard and communicate directly with each other, boosting secure inter-satellite communication. This increases responsiveness and efficiency, enabling satellites to adapt quickly to changing conditions, manage resources more effectively, and maintain operations even when ground contact is limited.

\end{itemize}

\subsubsection{Real-Time Autonomous Decision Making}

Space-based PKI systems enable secure, \textit{near real-time decision-making} by allowing autonomous satellites to authenticate and exchange critical data directly, \textit{minimizing latency}. This is crucial for time-sensitive tasks like disaster monitoring and collision avoidance, where rapid and reliable actions are essential. 

\begin{itemize}
   \item \textit{Reduced Latency:} Ground-based certificate validation involves latency, which is the round-trip time from a LEO satellite to data relay satellite(s) and then to the ground. In-space verification removes this dependency and reduces the latency, which is critical for time-sensitive operations like disaster monitoring.
   
   \item \textit{Near Real-Time Decision Making:} Autonomous satellites must make near real-time decisions. A key application is space traffic management (STM). As the orbital environment becomes increasingly complex, the risk of collisions with operational spacecraft, inactive satellites, or debris continues to grow \cite{NASA2021}. STM provides a coordinated approach to address these challenges by rapidly disseminating critical information about potential threats, enabling satellites to take swift, autonomous action \cite{STM2019}. Furthermore, coordination among satellites is essential, requiring the exchange of confidential maneuver information to ensure that avoidance actions, whether between active satellites or in response to non-maneuverable debris, effectively reduce risk rather than increase it. Space-based certificate validation ensures immediate authentication of alerts while maintaining the confidentiality of shared maneuver information.

\end{itemize}

\subsubsection{Scalability and Growth of Satellite Constellations}
\begin{itemize}
    \item \textit{Large-Scale Constellations:} With the deployment of mega-constellations like Starlink and OneWeb, the number of satellites in orbit is rapidly increasing. Managing certificate verification for thousands of satellites from the ground can be cumbersome. A space-based PKI system can streamline this process. Furthermore, in mega-constellations, inter-satellite communication is crucial for efficient operations. Space-based PKI systems ensure secure authentication, allowing satellites in mega-constellations to trust and reliably communicate with one another. These systems also scale efficiently, enabling the secure integration of new satellites into the network with minimal disruption.
    
    \item \textit{Dynamic Network Topologies:} Satellites are frequently reconfiguring their positions or switching roles within a constellation, resulting in dynamic network topologies. Space-based PKI allows for dynamic and flexible certificate management suited to these evolving configurations.

    \item \textit{Data Relay Networks:} In data relay systems, where data is passed between multiple satellites managed by different operators, space-based PKI systems ensures that each node in the network can authenticate and trust the other nodes.
\end{itemize}

\subsubsection{Global Coverage of Natural Phenomena and Disaster Areas}

\begin{itemize}
    \item  \textit{Cooperative and Timely Investigations of Natural Phenomena and Disaster Areas:} Spacecraft from different operators can exchange data and coordinate the study of phenomena and events near or on Earth through intercommunication capabilities. Impending disasters can be predicted in near real-time using this ad hoc networking, which facilitates rapid data collection and transfer to other satellites and ground stations  \cite{intercomm2001}. Satellites equipped with sensors that detect early warning signs can share this data with nearby satellites, enabling collaborative monitoring with complementary sensors for improved disaster assessment. For example, in the case of fire disasters, satellite-generated maps of dry areas and lightning strikes could be cross-referenced between satellites above the affected region to identify potential hotspots for fire outbreaks, enabling targeted monitoring of those regions. The data collected from disaster areas can be shared across all satellites, enhancing the timeliness and continuity needed to support ground emergency operations in addressing floods, fires, or other disasters. This collaborative approach ensures a more efficient and coordinated response to critical situations. Space-based PKI systems ensure secure, authenticated communication between satellites, which is crucial for sharing disaster-related data and supporting timely, coordinated emergency responses.
\end{itemize}

\subsubsection{Enhanced International Collaboration}
   \begin{itemize}

        \item \textit{ International Partnerships and Collaborative Missions:} Secure inter-satellite communication can significantly enhance international collaboration by creating more interconnected, shared, and efficient space networks that multiple nations can access and benefit from. By establishing secure links between satellites owned by different nations or organizations, they can pool resources, share data, and optimize the use of satellite capabilities. This reduces the cost burden on individual countries, particularly those with limited budgets for space missions. This also helps control the congestion of satellite traffic avoiding orbital crowding and managing limited orbital slots efficiently as space is a `limited space territory'. Space-based PKI systems support secure communication and data sharing among international partners and collaborative missions, ensuring that all parties can trust the authenticity of the exchanged information.
    \end{itemize}

\subsubsection{Support for Deep Space Missions}
       Deep space missions are not the focus of this paper; however, it is important to highlight the need for space-based PKI systems in these contexts. 
       
    \begin{itemize}
        \item \textit{Long-Duration Missions:} Missions to distant destinations such as Mars and beyond face significant communication delays due to the vast distances and the speed of light. This delay makes real-time coordination and support from Earth difficult. These missions must operate autonomously, particularly when managing critical systems like navigation, data processing, and security protocols. Space-based PKI systems can empower these missions to securely authenticate and verify communications and commands on their own, reducing the reliance on Earth-based systems.      
                    
        \item \textit{Interplanetary Communication:} Future interplanetary networks will require reliable, secure communication across vast distances between planets, moons, and spacecraft. A space-based PKI system across the solar system would be essential to maintain security and trust among various assets in these networks, providing real-time authentication, encryption, and secure communication channels.
       
    \end{itemize}
   
\subsubsection{Support for Emerging Use Cases in Space}

New use cases in space are constantly emerging, each requiring the adoption of space-based PKI systems to ensure secure, autonomous operations and communication, eliminating reliance on ground-based systems.

\begin{itemize}

    \item \textit{Space-based Data Centers:} The European Space Agency (ESA) has recently funded a project to explore the use of space-based data centers \cite{ESA-2024,ASCEND-2024}. Today, satellites usually collect data and send large amounts of raw information to Earth for processing. This approach is resource-intensive and can introduce delays. In contrast, in-orbit data centers process and store data directly in space. They can send back processed results or final insights either to the original satellite or to Earth, reducing the need to transmit large volumes of raw data \cite{ESA-2024,ASCEND-2024}. ESA has explored several deployment scenarios \cite{ESA-2024}: (i) two satellites in the same orbit, where one collects data and the other acts as a data center; (ii) a LEO satellite sending data to a data center in GEO; and (iii) exploration rovers transmitting data to a lunar lander that functions as a data center. These systems can support applications such as Earth observation, disaster management, deep-space missions, and inter-satellite analytics. There is also potential to offer these data center services commercially, creating new revenue streams \cite{ASCEND-2024}. However, since multiple operators may access shared computing resources, strong authentication is critical to prevent unauthorized access, data tampering, and misuse. As the number of satellites and space assets grows, authentication systems must scale efficiently. In this context, space-based PKI plays an important role by enabling secure authentication, real-time verification, and trusted interactions across the network.

    \item \textit{Federated Satellite Systems}: Federated Satellite Systems (FSS) \cite{GOLKAR2015230} envisions a space-based resource market where individual missions can act as suppliers or customers based on their needs, dynamically offering underutilized capabilities—such as bandwidth, processing power, or data storage—to the federation. FSS requires verification of both suppliers and customers \cite{VONMAURICH2018} in space. In FSS, a space-based PKI model enables each participant in the federated satellite system to verify the identity of all registered spacecraft.

    \item \textit{Software-based Virtual Missions:} Software-based virtual space missions \cite{Loft-2024} refer to the use of existing satellites to achieve new and dynamic objectives, enhancing operational flexibility and efficiency while minimizing the costs associated with deploying additional satellites. Instead of launching new satellites, these missions allow operators to reconfigure and re-purpose in-orbit satellites for additional tasks or entirely new objectives. Virtual missions can be deployed quickly and cost-effectively, bypassing the logistical challenges of satellite launches. For instance, satellites can be dynamically configured for urgent, short-term tasks such as disaster monitoring or scientific observation, enabling real-time responses to critical needs. Consequently, a single satellite may undertake multiple missions either sequentially or concurrently. These virtual missions, originating from different entities or operators with distinct PKIs and varied communication requirements, highlight the critical need for a robust space-based PKI system.

\end{itemize}


\subsection{Implementation Challenges of Space-Based PKI}

While space-based PKI offers the advantage of in-orbit certificate issuance and validation, several key challenges must be addressed for effective deployment in dynamic, resource-constrained space environments, including:

\begin{enumerate}
     
    \item \textbf{Heterogeneous PKI systems.} With the growing diversity of PKI systems used by space agencies, commercial operators, and international partners, ensuring scalable and robust \textit{interoperability} is critical. Effective interoperability requires a unified framework that allows entities across multiple organizations to validate and trust certificates issued by different CAs, supporting secure and reliable communication.
    
    \item \textbf{Certificate renewal, revocation, and updates.} Regularly updating certificates renewal and revocation information presents significant challenges in space-based environment. Ensuring that all nodes, including satellites and ground stations, have the most current certificate status is essential for maintaining security and preventing unauthorized access. The process must account for \textit{limited communication windows}, \textit{dynamic orbital paths}, and \textit{intermittent connectivity}.   

    \item \textbf{Real-time authentication.} \textit{Connectivity gaps} can disrupt data exchange and authentication, delaying certificate validation and integrity checks. In space-based systems, where continuous communication is often infeasible, these gaps hinder real-time authentication needed for secure operations.

    \item \textbf{Bandwidth limitations.} Bandwidth is a critical resource in space communication, directly affecting \textit{cost}, \textit{energy consumption}, and \textit{frequency allocation}. Uplink and downlink channels are highly asymmetric, with uplinks typically offering much lower bandwidth than downlinks. PKI services must have lightweight protocols to prevent overloading the space link.
    
    \item \textbf{On-board resource constraints.} Satellites have limited \textit{computational power}, \textit{storage capacity}, and \textit{energy} compared to ground stations. Hence, efficient cryptographic algorithms and optimized software are essential to handle PKI operations.

\end{enumerate}

The first two challenges are commonly encountered by ground-based PKI systems; however, they become more complex when combined with additional challenges unique to space environments, such as \textit{intermittent connectivity}, \textit{dynamic orbital paths}, \textit{extended communication delays}, and \textit{restricted resources} available on satellites.


\section{Preliminaries}
\label{background}

\subsection{Public Key Infrastructure and Interoperability} 

The main components of a \textit{\textbf{Public Key Infrastructure (PKI)}} include the \textit{Registration Authority (RA)}, \textit{Certificate Authority (CA)}, \textit{Certificate Repository (CR)}, \textit{Certificate Policy (CP)}, \textit{Certificate Validation Process (CVP)}, and \textit{Relying Party (RP)} \cite{rfc3647,rfc4158,rfc6960}. These components are organized in a PKI architecture, which can be broadly classified into \textit{single CA}, \textit{hierarchical}, and \textit{mesh} architectures. A detailed description of PKI components and architectures is provided in the Appendix~\ref{apndx:PKI} for readers seeking additional background. The \textit{\textbf{interoperability}} of PKI systems in multi-operator networks allows different operators' CAs and users to trust and validate each other's certificates seamlessly. This is essential for establishing trust and secure communication across diverse entities. The interoperability mechanisms such as \textit{Bridge Certification Authority} and \textit{Validation Authority} enable trust relationships and certificate validation in complex, distributed, and heterogeneous environments.

\subsubsection{Bridge Certification Authority} 

\begin{figure}[ht]
    \centering
    \includegraphics[width=0.90\textwidth]{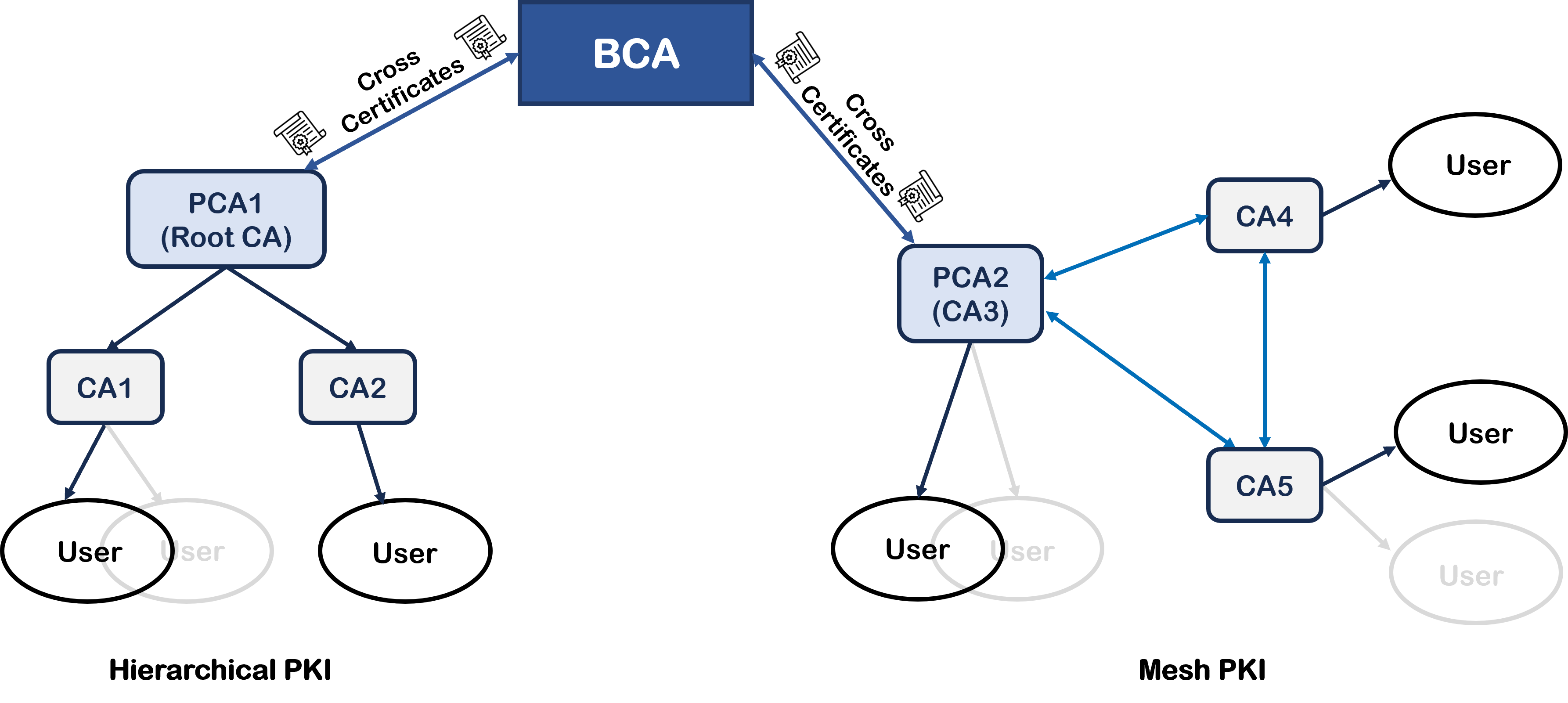}
    \caption{Bridge Certification Authority (BCA) connecting different PKI architectures}
    \label{fig:BCApki}
\end{figure}

The Bridge Certification Authority (BCA) trust model \cite{BCA} was designed to ensure interoperability across heterogeneous PKI environments, as illustrated in Figure~\ref{fig:BCApki}. This model connects different PKI architectures by introducing the Bridge CA, which facilitates relationships between various PKIs. A CA within a PKI, called \textit{Principal Certification Authority (PCA)}, is designated to establish a trust relationship with the BCA. In hierarchical PKIs, BCA establishes relationships with the root CA (e.g., Root CA in Figure~\ref{fig:BCApki}). In mesh PKIs, BCA establishes a relationship with any one CA (e.g., CA3 in Figure~\ref{fig:BCApki}), and all of the communities in the mesh PKI become part of the resulting larger PKI. Each trust relationship is represented by a pair of certificates -- one issued by the BCA to a PCA and the other issued by the PCA to the BCA, thus forming a “bridge of trust” between users from different PKIs. RP establishes a certificate path to a target CA by initiating from its own trusted root authority and sequentially validating the certificate issued by the root to the BCA, following the chain of trust. Unlike a CA in a mesh PKI, the BCA does not issue certificates to end users, and unlike a Root CA in a hierarchical PKI, it does not serve as a trust anchor. The BCA takes care of the following main responsibilities:
     
    \begin{enumerate}
        \item \textit{Cross-Certification.} Establishes trust with participating CAs through cross-certificates.
     
        \item \textit{Policy Mapping.} Provides policy mappings to align certificate policies across different CAs, translating differences in trust models or rules
                
 
    \end{enumerate}

BCAs are typically deployed following industry standards such as X.509 certificate format and certificate path validation (RFC 5280 \cite{rfc5280}), and cross-certification (RFC 3379 \cite{rfc3379}). These standards ensure uniform compliance among participating CAs, allowing seamless interoperability across systems. A real-world example is the U.S. Federal Bridge CA (FBCA) \cite{WEBurr1998,ALTERMAN2001} for secure communication across federal agencies.

\subsubsection{Validation Authority}

\begin{figure}[ht]
    \centering
    \includegraphics[width=0.9\textwidth]{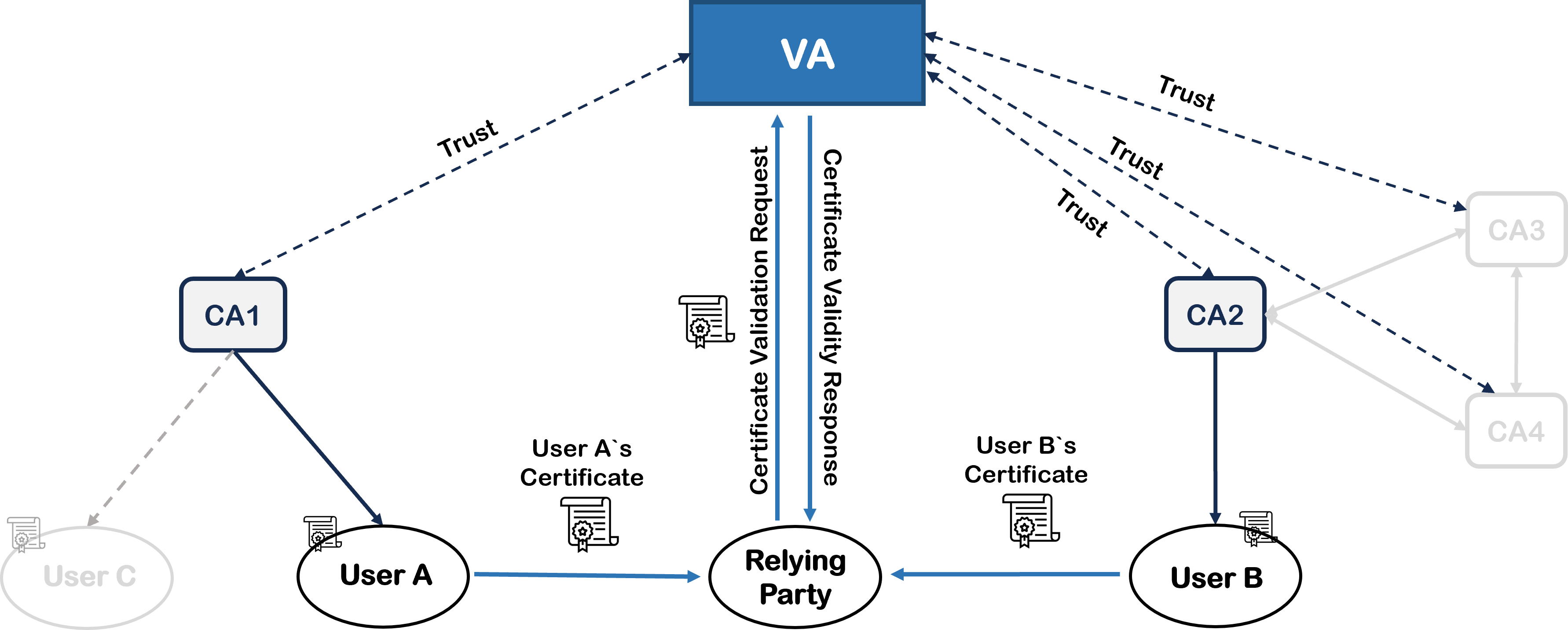}
    \caption{Independent Validation Authority (VA)}
    \label{fig:VApki}
\end{figure}

In the Validation Authority (VA) trust model \cite{Jon2006}, an independent, trusted third party serves as the trust anchor for certificate verification on behalf of RPs (Figure~\ref{fig:VApki}). Unlike a BCA, the VA does not issue any certificates. Once trusted, it provides a single point of trust, allowing recipients to trust all CAs managed by the VA without knowing the broader trust structure. VA maintains its own trust list of verifiable CAs, so certificate path discovery and validation are unnecessary, though they may still be used internally. With VA, a BCA becomes redundant, as VA alone can ensure interoperability \cite{Jon2006}, provide dynamic certificate validation, and simplify trust management. The VA takes care of the following tasks:
 
    \begin{enumerate}
        \item \textit{Certificate Status Validation.} Validates the current status of a certificate on behalf of a RP.
    \end{enumerate}

VA operates in accordance with international standards and guidelines for PKI, such as X.509 Certificates format (RFC 5280 \cite{rfc5280}) and OCSP (RFC 6960 \cite{rfc6960}). 

\section{Space-Based Public Key Systems and Infrastructure}

\label{PKI}

As of today, there is no dedicated PKI certificate authority or certificate validation service physically deployed in space. While the concept of a space-based PKI system is yet to be realized, its implementation will become essential as space operations grow increasingly sophisticated and autonomous. As satellite networks manage complex tasks without constant ground-based control, the demand for space-based PKI servers will rise. Accordingly, two PKI deployment approaches are proposed in this work. The first, \textit{space-ground integrated PKI (iPKI)}, enables participating space objects in LEO, MEO, and even GEO to get the certificate validation done directly in space, regardless of the originating PKI domain on ground. This proposal offers an incremental improvement over existing space-based PKI methods. Building on existing approaches allows for a gradual, incremental improvement without disrupting current systems. While effective, issuing certificates on the ground and verifying them in space can impose a considerable burden on uplink and downlink channels. However, the uplink, which carries only a small amount of traffic (mainly commands sent to the satellites), could be enhanced to accommodate the PKI. Given the low traffic on the uplink, consisting primarily of commands, implementing the iPKI is feasible. The second approach, a \textit{space-based PKI (SpcPKI)}, implements a fully in-space PKI, offering greater autonomy, scalability, and reduced operational complexity. The SpcPKI enables participating space objects in LEO, MEO, and GEO to utilize complete PKI functionality all managed in space by space-based CAs. First approach extends the existing ground-based PKI to support certificate verification in space, while the second adopts a clean-slate, fully space-resident PKI. These are alternative deployment strategies, each capable of supporting the same missions, with the choice driven by deployment and cost considerations rather than mission scope.


\subsection{Space-Ground Integrated PKI}

The space-ground integrated PKI (iPKI) leverages BCA and VA, which perform complementary roles. While VA services often make BCAs unnecessary in ground networks \cite{Jon2006}, relying solely on VA is less practical in space due to transmission delays, limited communication windows, and intermittent connectivity. In iPKI, the \textit{ground-based BCA} establishes trust among diverse PKI domains, particularly valuable in large-scale, multinational collaborations among different space agencies and spacecraft. PKI interoperability is recognized by CCSDS as essential for secure cross-support and coordination among agencies and partners in space missions \cite{CCSDS357.1-O-1}. The \textit{space-based VA}, in turn, enables near real-time certificate validation across these domains, occurring in space, ensuring secure space communication. By tailoring their responsibilities within the iPKI to the unique demands of space operations, the BCA and VA provide a scalable, interoperable, and resilient framework. This framework continuously maintains trust relationships and ensures timely certificate validation and compliance checks in space.

\subsubsection{Primary Entities in iPKI}

The entities in iPKI include Space-based RP (Spc-RP), Space-based VA (Spc-VA), Ground-based CA (Grd-CA), Ground-based BCA (Grd-BCA), and Ground-based PCA (Grd-PCA). The roles of Spc-RP and Grd-CA are the same as in terrestrial PKI systems. The Grd-PCA, however, performs some additional tasks. Moreover, a Grd-BCA and a Spc-VA play distinct but complementary roles that are adapted to facilitate trust, certificate validation, and secure communication among different PKI domains. The roles of these entities in iPKI are described below.

\begin{itemize}

    \item \textit{\textbf{Spc-RP}} represents a space entity, e.g., a satellite, that needs to validate a certificate received from other space entity or ground for authentication or confidentiality.
    
    \item \textit{\textbf{Spc-VA}}, deployed in space, is mainly responsible for the \textit{real-time validation of certificates} in space. The Spc-VA is managed under iPKI service contract and takes care of the following responsibilities:

    \begin{itemize}
        \item Unlike traditional VA, establishes and maintains a \textit{certificate repository (VA-Repository)} in space, used to store certificates, certificate policies, certificate policy mappings, and the certificate revocation information.

        \item On receiving a certificate validation request from a Spc-RP, checks the \textit{validity} status of the certificate, the \textit{compliance} of the certificate with the policies and trust rules established by the Grd-BCA, as well as the key usage restrictions, and performs certificate \textit{path discovery and validation}. 
    \end{itemize}

    \item \textit{\textbf{Grd-CA}}, on the ground, issues, renews, and revokes certificates for end users such as satellites, ground stations, or other space entities.
        
    \item \textit{\textbf{Grd-BCA}}, on ground, serves as a trusted root for space systems, establishing trust among disparate ground-based space PKI systems and between ground-based PKI systems and the Spc-VA, and is managed under iPKI service contract. The role of Grd-BCA is also modified in iPKI as follows:

    \begin{itemize}
        \item Like traditional BCA, the Grd-BCA establishes \textit{cross-certification} with the Grd-PCAs of the space agencies. Additionally, it issues a certificate to Spc-VA, establishing a trusted bridge between the ground-based PKIs and the Spc-VA for later certificate validation purposes.
    
        \item  Additionally, it performs certificate \textit{policy mappings} and \textit{uploads the mapping information} to the repository maintained by the Spc-VA (\textit{VA-Repository}). 

        \item \textit{Uploads certificate revocation information} to the \textit{VA-Repository} including revocation data provided by the Grd-PCAs regarding their issued certificates, as well as the revocation information of the Grd-PCAs themselves. 

    \end{itemize}

    \item \textit{\textbf{Grd-PCA}}, within each participating ground-based PKI, establishes a trust relationship with the Grd-BCA through \textit{cross-certification}. The Grd-BCA issues one certificate to the Grd-PCA, and the Grd-PCA issues one to the Grd-BCA, in accordance with the defined policy guidelines, as is done in terrestrial networks.

    \begin{itemize}
        \item In iPKI, the Grd-PCA is also responsible for supplying the \textit{intermediate CA certificates}  for the \textit{VA-Repository} that form the complete certification paths from any end user up to the Grd-PCA certificate issued by the Grd-BCA. It is also tasked with providing the \textit{certificate policies}.

        \item Additionally, it provides the up-to-date \textit{revocation status} of the end user certificates as well as intermediate CA certificates issued under its PKI to Grd-BCA through CRLs or other mechanisms.

    \end{itemize}

\end{itemize}

\subsubsection{Functional Description of iPKI}

Building on the defined entities, this subsection describes the functional operations of the proposed iPKI to enable secure and efficient PKI operations on ground and in space.

\medskip

\noindent \textbf{A. Trust Establishment and Interoperability} 

\begin{itemize}
    \item \textit{\textbf{Initial Setup.}} The Spc-VA is an always available trusted service working in collaboration with Grd-BCA. The Grd-BCA, as a self-certified CA, issues a certificate to Spc-VA. The Spc-VA also obtains the certificate of Grd-BCA. The \textit{VA-Repository} is set up in space to store certificates, certificate policies, policy mappings, certificate revocation information, and other essential data required for authenticating the chain of trust. The Spc-VA accesses this repository to process certificate validation requests sent by Spc-RPs effectively and efficiently.
    
\end{itemize}

\begin{itemize}
    
    \item \textit{\textbf{Cross-Certification.}} 
    Any satellites or space agency with a ground-based PKI that needs to participate in secure space communications registers with the iPKI. After authentication, registration, and service agreement, the Grd-BCA and the Grd-PCA of participating ground-based PKI cross-certify each other, forming an integrated multi-entity PKI (Figure~\ref{fig:BCpki}). This makes the Grd-BCA a trusted bridge for all participating PKI entities, including satellites as certificate holders and relying parties. Cross-certification enables secure communication both among participating space agencies on the ground and between space assets in space, thereby eliminating the need for a separate BCA dedicated to ground communication. At this stage, each Grd-PCA also gains access to the Grd-BCA certificate, which is later used by the Spc-RPs to verify the certificate validation response from the Spc-VA.

    \begin{figure}[ht]
        \centering
        \includegraphics[width=0.9\textwidth]{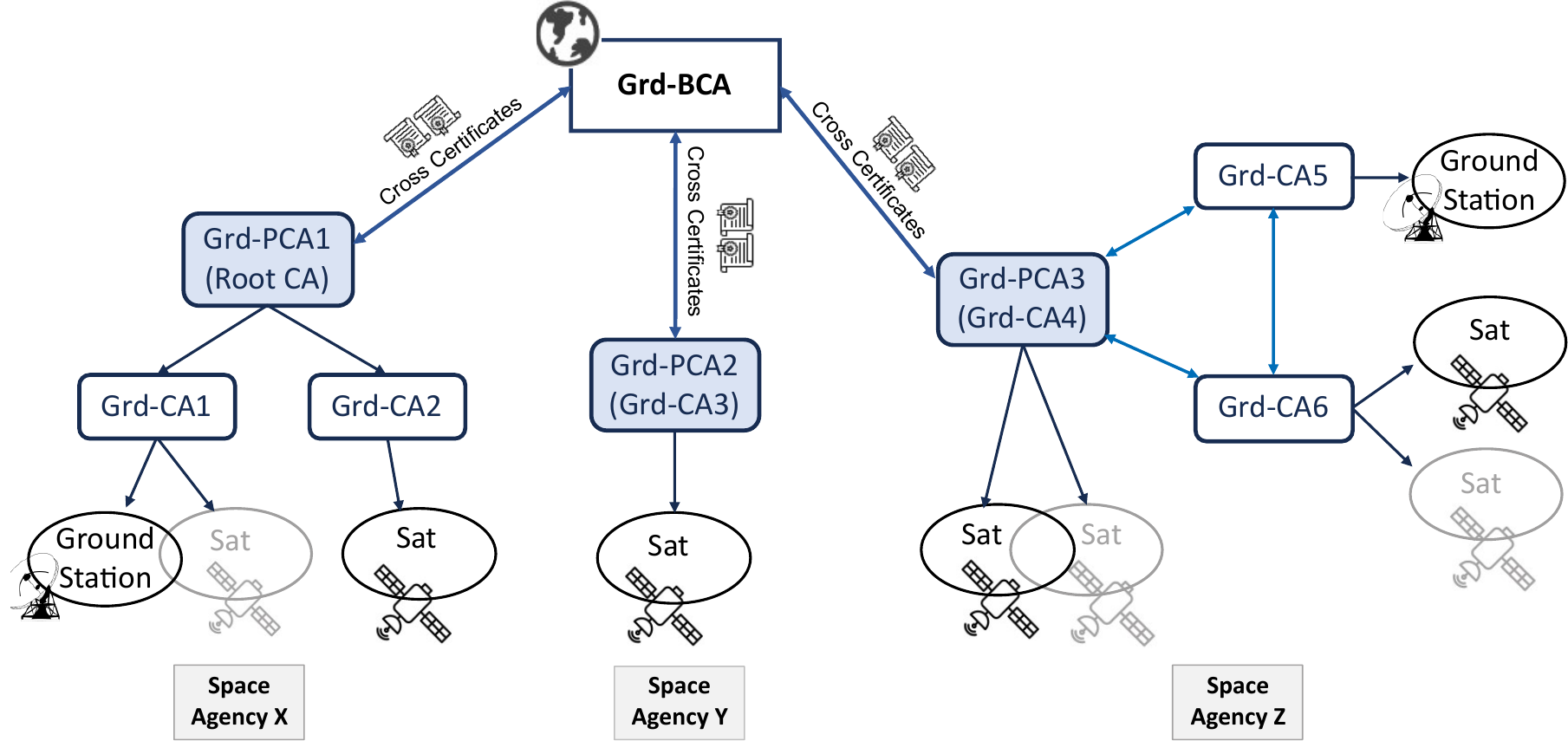}
        \caption{Grd-BCA connecting disparate ground-based PKIs via cross-certification}
        \label{fig:BCpki}
    \end{figure}

\end{itemize}



\begin{itemize}
    \item \textit{\textbf{Certificates and Policy Mapping Information.}} The Grd-BCA performs necessary policy mappings to align certificate policies across different PKI domains. Once cross-certificates are issued, the Grd-BCA uplinks Grd-PCAs's certificates to \textit{VA-Repository} together with the policy mapping information. Whenever there is a change, the Grd-BCA updates that change to \textit{VA-Repository}.
    
    \item \textit{\textbf{Intermediate Certificates.}} The Grd-BCA also uploads the trust chain certificates (from Grd-PCA to the Grd-CA certificates) and corresponding policies received from Grd-PCAs and keeps updating this information whenever there is a change. 
\end{itemize}

Note that there are not many space agencies around the world, therefore, only a limited number of diverse PKI systems may be involved.

\medskip

\noindent \textbf{B. Certificate Renewal and Revocation} 

\begin{itemize}
    \item \textit{\textbf{Certificate Renewal.}} Depending on the mission’s nature and lifecycle, satellites may operate for several years. To balance the need for security and avoid frequent certificate renewals, space missions are expected to use certificates valid for longer periods of one to three years \cite{CCSDS357.1-O-1}. These certificates can be renewed as needed. Cross-certificates and trust chain certificates may also be renewed, if needed. The Grd-BCA publishes the renewed certificates (if any) to the \textit{VA-Repository}.

    \item \textit{\textbf{Intermediate CA and End User Certificate Revocation.}} The Grd-PCA sends up-to-date revocation information of intermediate CAs and end users certificates to Grd-BCA as soon as there is any update. Grd-BCA disseminates this information to \textit{VA-Repository} through CRLs or any other mechanism as soon as it receives it.
    
    \item \textit{\textbf{Cross-Certificate Revocation.}} Grd-PCA's certificate may also be revoked due to misuse, private key compromise, or any other reason. Grd-BCA publishes Grd-PCA's certificate revocation information to the \textit{VA-Repository} as soon as possible. 
\end{itemize}

\smallskip

\noindent \textbf{C. Near Real-Time Certificate Validation}  

\smallskip

\begin{figure}[ht]
        \centering
        \includegraphics[scale=0.5]{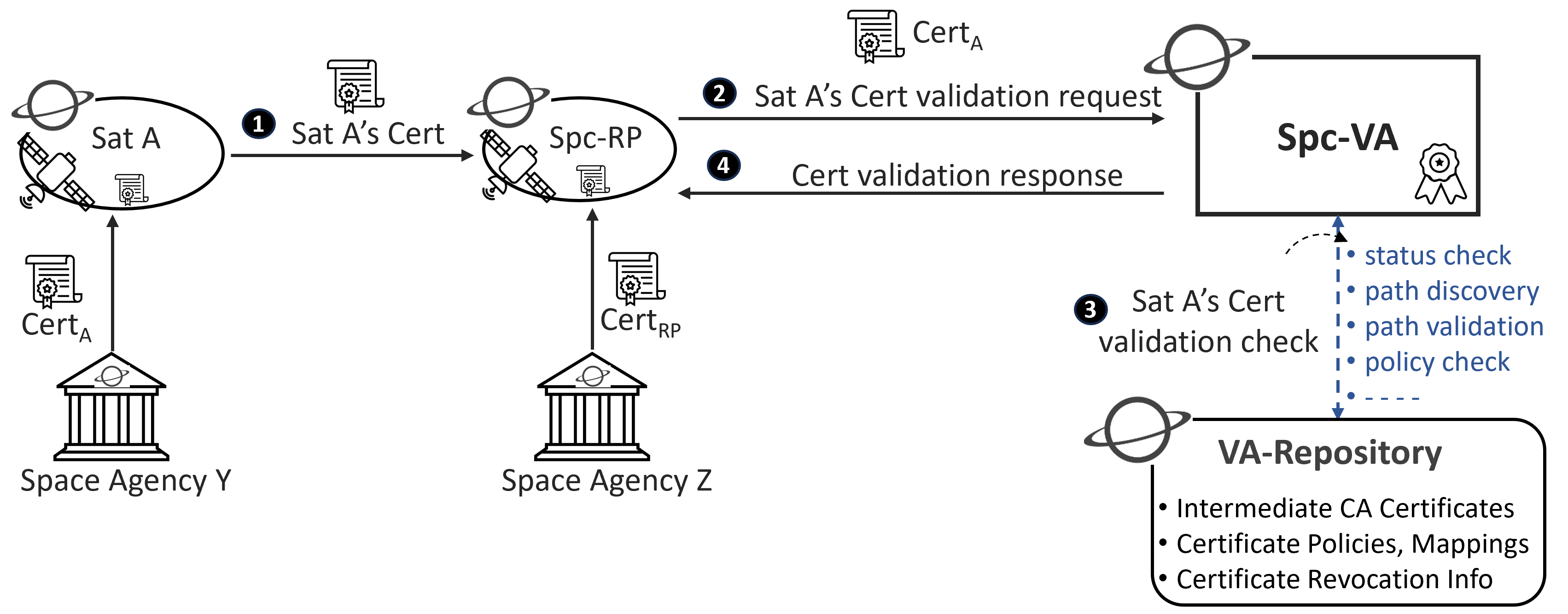}
        \caption{Near-real-time validation of certificates in iPKI}
        \label{fig:Cert_Valid_iPKI}
    \end{figure}

The Spc-VA operating in space validates a certificate on behalf of Spc-RP. The following steps outline the process (see Figure~\ref{fig:Cert_Valid_iPKI}):

\begin{itemize}
    \item \textit{\textbf{Certificate Validation Request.}} When a Spc-RP in space receives a certificate from a certificate holder, e.g., Sat A, particularly belonging to a different PKI system, Spc-RP sends the certificate validation request, along with Sat A's certificate $Cert_{A}$, to the nearest Spc-VA. This request is signed using Spc-RP's private key associated with a certificate $Cert_{RP}$ that chains up to the Grd-PCA's certificate, obtained from the Grd-BCA, as its root of trust. 

    \item \textit{\textbf{Certificate Validation Check.}} The Spc-VA accesses \textit{VA-Repository} to perform the certificate validation as follows:

    \begin{enumerate}

        \item \textit{Certificate Path Discovery \& Validation}
        \begin{itemize}
            \item \textit{Path Discovery.} Builds certificate path that links the Sat A's certificate to a Grd-PCA, a trust anchor for Sat A (Figure~\ref{fig:Cert_Path}). The path always begins with the self-signed root certificate of Grd-BCA. Next certificate in this path is the certificate issued by Grd-BCA to Grd-PCA of Sat A. After that will be the intermediate CA certificates (if any). Last certificate in this path will be Sat A's certificate signed by Grd-CA. Guidance and recommendations for building X.509 public-key certification paths are provided in \cite{rfc4158}.
            \item  \textit{Validity Period.} Checks the validity period of each certificate in the path.
            \item \textit{Path Validation.} Verifies the signatures on each certificate in the path 
        \end{itemize}  
                \item \textit{Certificate Revocation Check}
        \begin{itemize}

            \item \textit{Revocation Status.} Checks the certificate revocation status of each certificate in the path using the most up-to-date revocation information published by Grd-BCA to \textit{VA-Repository}.
        \end{itemize}           
        
        \item\textit{Policy Mapping and Compliance Verification}
        \begin{itemize}
            \item \textit{Policy Compliance Check.} Validates that the certificate policies align with the relying party’s requirements. 
            
            \item \textit{Policy Mapping Check.}  The Spc-VA uses the policy mapping information issued by the Grd-BCA to validate whether a certificate from Sat A's PKI domain is compliant with the policies of Spc-RP's domain.
        \end{itemize}
        
    \end{enumerate}

    \begin{figure}[ht]
            \centering
            \includegraphics[width=0.8\textwidth]{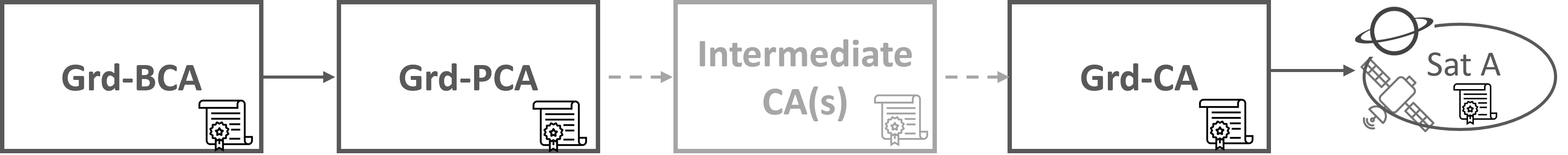}
            \caption{Certification Path}
            \label{fig:Cert_Path}
    \end{figure} 

    \item \textit{\textbf{Certificate Validation Response.}} The Spc-VA either confirms or denies the trustworthiness of the certificate based on its validation checks and the information provided by the Grd-BCA and sends the certificate validation response back to Spc-RP. For a signed response from Spc-VA, the Spc-RP can verify the Spc-VA certificate using Grd-BCA's certificate.
    
\end{itemize}

\subsection{Dedicated Space-Based PKI}

In dedicated space-based PKI (SpcPKI), a CA on the ground helps establish trust among entities from diverse PKI domains that wish to use the services provided by SpcPKI. Meanwhile, the CA operating in space handles certificate issuance, renewals, revocations, and real-time validation of certificates issued by SpcPKI in space. A BCA, for interoperability, is not required, as all entities will operate under a unified PKI once deployed in space. The RP can verify an end-user certificate independently, without relying on the CAs. However, if needed, the CA can also verify the certificate on behalf of the RP. Moreover, VAs can be deployed separately, as in iPKI, allowing more VAs than CAs to be used to improve certificate verification coverage. The roles and responsibilities of the ground-based CA and space-based CA are tailored to meet the unique demands of space operations. 

\subsubsection{Primary Entities in SpcPKI}

The entities in SpcPKI include Space-based RP (Spc-RP), Space-based CA (Spc-CA), and Ground-based CA (Grd-CA). The roles of these entities in SpcPKI are described below.

\begin{itemize}

    \item \textit{\textbf{Spc-RP}} retains the same role as in iPKI.
    
    \item \textit{\textbf{Spc-CA}} is responsible for certificate lifecycle management and, if required, performing \textit{real-time validation of certificates} in space. It is managed under the SpcPKI service contract and may be structured hierarchically. It takes care of the following responsibilities:

    \begin{itemize}
        
        \item Issues new certificates to end users, renews expiring ones, and revokes certificates that are malicious or no longer needed.
        
        \item Like the CA in terrestrial networks, establishes and maintains a \textit{certificate repository (CA-Repository)}, used to store certificates, certificate policies, certificate policy mappings, and certificate revocation information.

        \item Validates the certificate received from Spc-RP as is done by Spc-VA in iPKI. 
        
    \end{itemize}
    
    \item \textit{\textbf{Grd-CA}}, on ground, serves as the initial trust anchor for space systems, establishes trust between end users and the space CA, and is managed under a PKI service contract. It issues certificates to end users, such as satellites or any other entity involved in space communication, after identity verification. There may be more than one Grd-CA deployed, optionally structured hierarchically.
\end{itemize}

\subsubsection{Functional Description of SpcPKI}

This subsection describes how the defined entities in SpcPKI function securely and efficiently on the ground and in space.

\medskip

\noindent \textbf{A. Trust Establishment and Interoperability} 

\begin{itemize}
    \item \textit{\textbf{Initial Setup.}} The Spc-CA is a trusted, always-online service that operates in collaboration with the equally trusted and available Grd-CA. The Grd-CA, as a self-certified CA, issues a certificate to Spc-CA who also obtains the Grd-CA's certificate. The \textit{CA-Repository} is set up in space. 
\end{itemize}

\begin{figure}[ht]
    \centering
    \includegraphics[scale = 0.52]{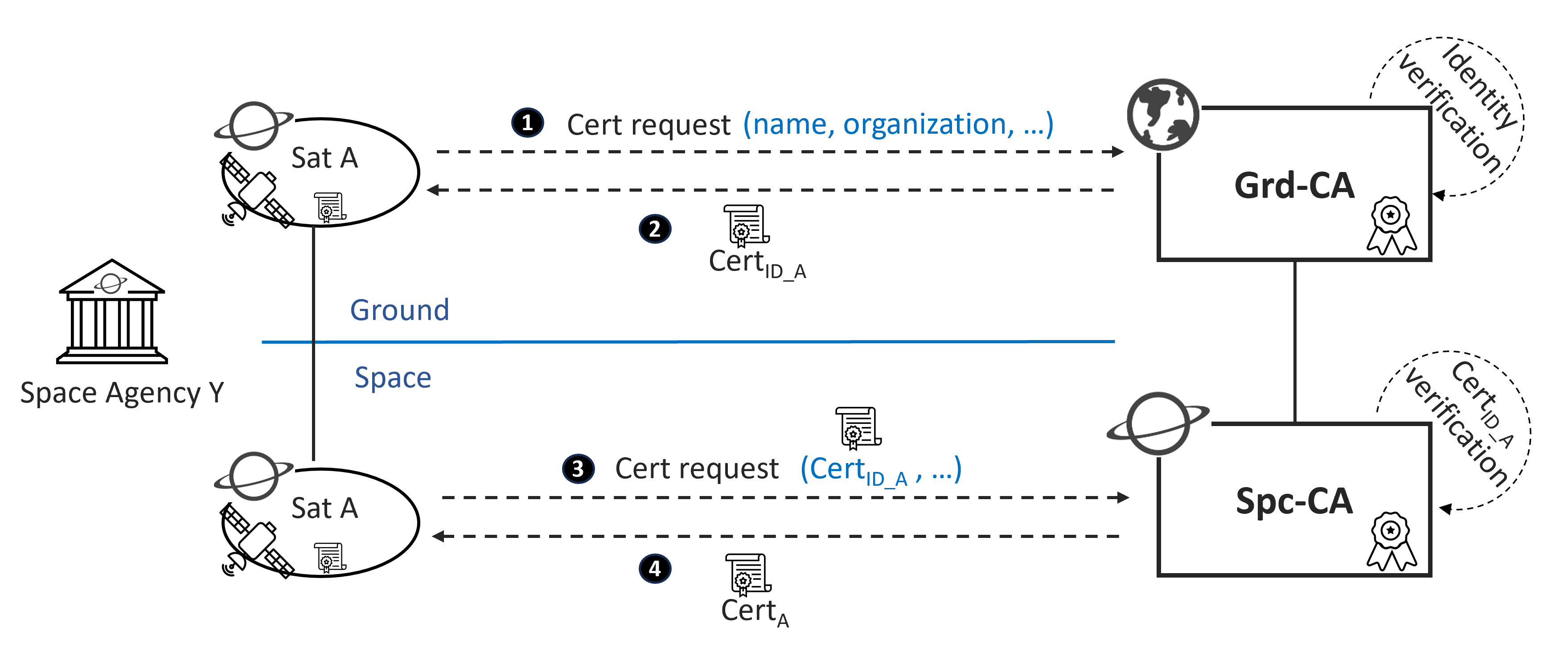}
    \caption{Certificate Issuance in SpcPKI}
    \label{fig:Cert_issue_SpcPKI}
\end{figure}

\begin{itemize}
    
    \item \textit{\textbf{Certification on ground.}} Any ground-based PKI system wishing to participate in secure space communications must first register with the SpcPKI. Grd-CA verifies the identity information (e.g., name, organization, domain) either  online or offline. After authentication, registration, and service agreement, the Grd-CA issues a certificate (e.g., $Cert_{ID\_A}$) to the end user to prove their authenticity to the Spc-CA. At this stage, the end user also obtains the Spc-CA certificate, which is later used to authenticate the Spc-CA in space.

    \item \textit{\textbf{Certification in Space.}} While in space, an end user (e.g., Sat A) with an initial certificate $Cert_{ID\_A}$ authenticates itself to Spc-CA. After successful authentication, Spc-CA issues a new certificate (e.g., $Cert_{A}$) to satellite for communication with space entities both inside and outside its own agency or organization (Figure~\ref{fig:Cert_issue_SpcPKI}). Spc-CA uploads certificates together with policy and mapping information to \textit{CA-Repository}.
\end{itemize}


Note that there are not many space agencies around the world, therefore, only a limited number of PKI systems may be involved.

\medskip

\noindent \textbf{B. Certificate Renewal and Revocation}  

\begin{itemize}
    \item \textit{\textbf{Certificate Renewal.}} The Spc-CA may renew a user certificate user upon request. The renewed certificate is uploaded to the \textit{CA-Repository}.
    
    \item \textit{\textbf{Certificate Revocation.}} The Spc-CA immediately updates certificate revocation information to the \textit{CA-Repository}. 

\end{itemize}

\noindent \textbf{C. Near Real-Time Certificate Validation} 

\smallskip

\begin{figure}[ht]
        \centering
        \includegraphics[width=0.85\textwidth]{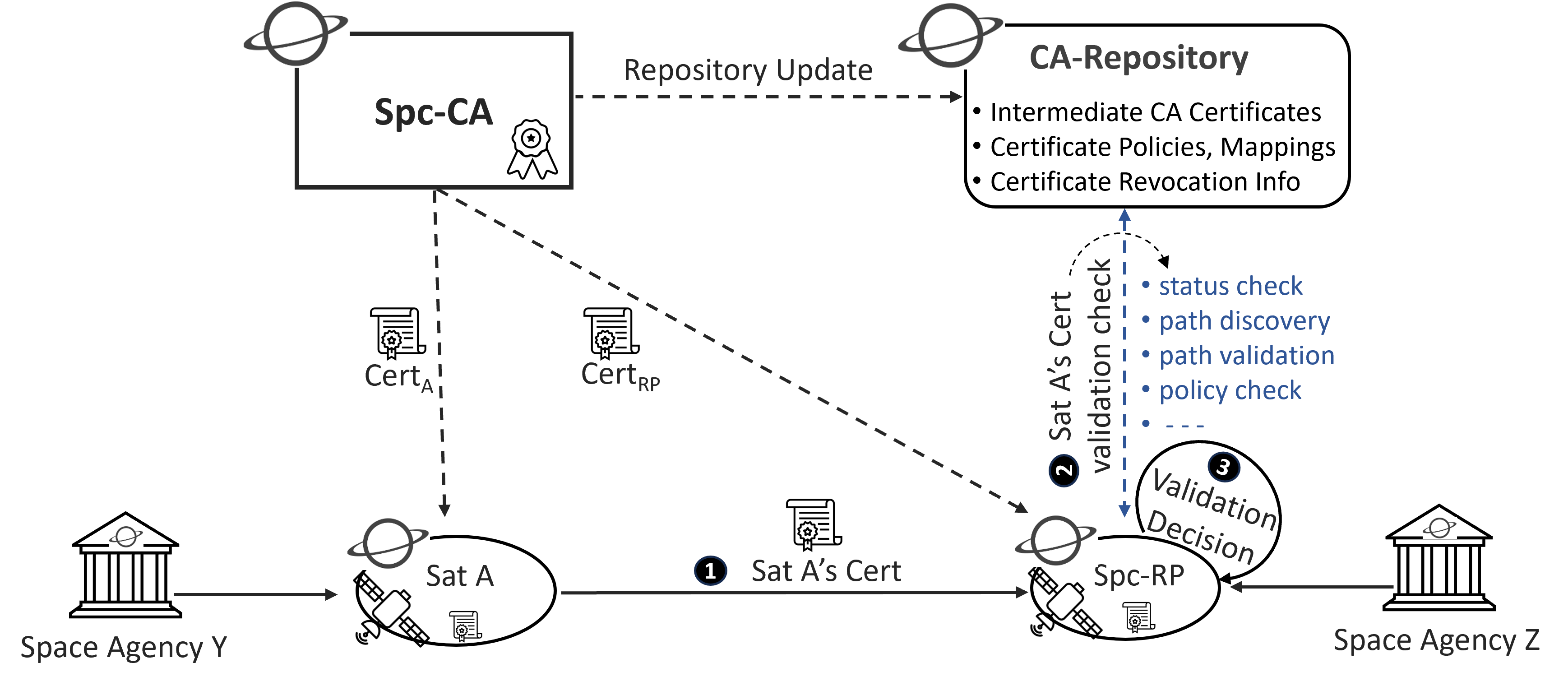}
        \caption{Near-Real-Time validation of certificates in SpcPKI}
        \label{fig:Cert_Valid_SpcPKI}
    \end{figure}

The Spc-CA or Spc-VA may validate a certificate on behalf of Spc-RP, following the same procedure as described earlier for iPKI (Figure~\ref{fig:Cert_Valid_iPKI}). However, in SpcPKI, the Spc-RP can also verify certificate directly, which is more suitable for larger, resource-rich satellites. The Spc-RP, on receiving a certificate from a satellite (e.g., Sat A), performs the certificate validation check as follows (Figure~\ref{fig:Cert_Valid_SpcPKI}):

\begin{itemize}
    \item \textit{\textbf{Certificate Path Discovery \& Validation.}} The Spc-RP constructs certificate path from the root of SpcPKI up to Sat A and validates each certificate. The certificate chain may reside in Spc-RP's local trusted storage, be provided by Sat A with the signed request, or be retrieved from \textit{CA-Repository} via Spc-CA.
    
   \item \textit{\textbf{Certificate Revocation Check.}} This can be handled in several ways. 
    \begin{itemize}
         \item Using CRLs stored on Spc-RP, sent by the SpcPKI to all space entities whenever there is a change.
        \item Querying a nearby Spc-CA within the SpcPKI system for the latest CRL.
        \item Following an adapted OCSP-style query--response or stapling model for space.
    \end{itemize}    

    \item \textit{\textbf{Policy Compliance Check.}} Checks compliance with applicable policies.

    \item \textit{\textbf{Certificate Validation Decision.}} The Spc-RP either confirms or denies the trustworthiness of the certificate based on its validation results.
\end{itemize}

However, given the constraints of space communication, existing mechanisms such as OCSP may need to be adapted for space environments, or entirely new lightweight protocols may need to be developed to support efficient certificate status checking.

\subsection{Deployment in Space: Design Considerations}

The technical implementation of space-based PKI systems is not the primary focus of this paper; however, key insights are provided for space-based entities and operations.

\medskip

\textbf{\textit{Spc-CA and Spc-VA Architectures.}} The functionality of the Spc-CA and Spc-VA must remain continuously accessible to all participating space objects. To support this, these entities should use a distributed architecture with sufficient redundancy to ensure high availability and fault tolerance despite satellite failures, communication disruptions, and other space-specific challenges. One way is to deploy multiple replicas of the Spc-CA and Spc-VA at different orbital locations, within the same orbit and across multiple orbits, so that if one instance becomes inaccessible due to orbital constraints, others can take over its role. Additionally, a hierarchical architecture for Spc-CAs can be considered, where a root Spc-CA oversees several subordinate Spc-CAs deployed across different orbital regions. This approach will enhance scalability, simplify trust management, and allow for localized certificate issuance and validation within orbital zones. The number and configuration of replicas or tiers will depend on factors such as the number of users in space, the orbital parameters (e.g., altitude, inclination, eccentricity), and operational requirements. These architectural considerations are particularly important, as the end certificate users, Spc-CA, Spc-VA, and Spc-RPs may all be moving in distinct orbits, introducing relative velocities that affect communication windows and system coverage.

\textbf{\textit{Certificate Repository.}} Repositories often require high availability to ensure consistent and reliable access to data. The \textit{Certificate Repository} for storing certificates and related information can be set up in multiple ways such as database storage, LDAP directory, etc. Similarly, it can be configured in different ways. One approach is a logically centralized repository, where PKI entities use separate databases that are networked together and appear as a unified system to users despite physical separation. This setup is ideal for space systems involving multiple trust domains and geographically dispersed entities. However, it may face challenges due to intermittent connectivity. Another option is a centralized repository with backups to a distributed system, combining both physical and logically distributed methods. This approach will enhance security and resilience, making it suitable for space agency networks or mission-critical infrastructures. However, this approach requires tight synchronization between repositories and optimized updates of revocation and auxiliary data to improve efficiency and reliability.

\begin{figure}[ht]
    \centering
    \includegraphics[width=0.5\textwidth]{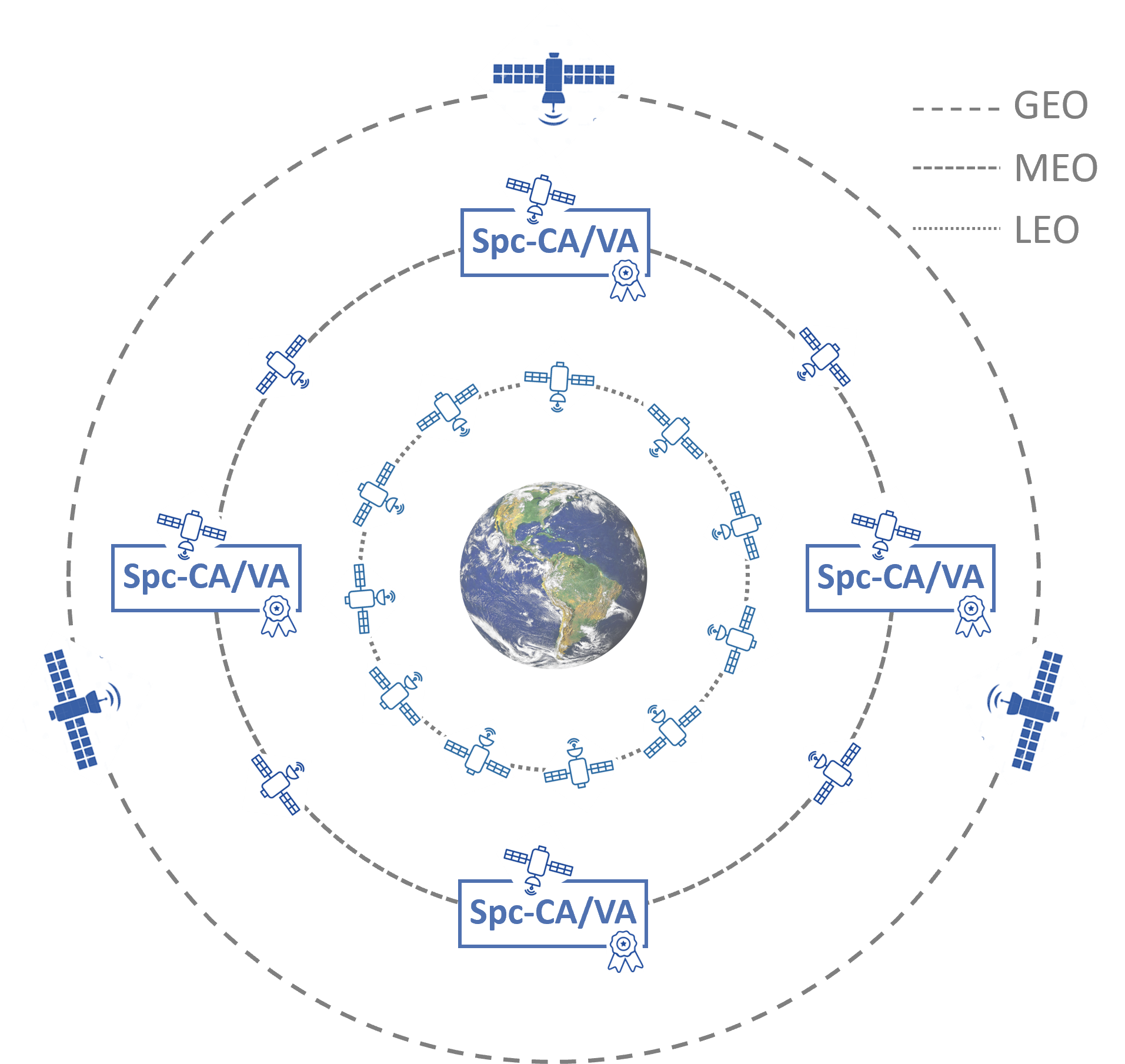}
        \caption{Spc-CA/VA in MEO}
        \label{fig:CA-VA_MEO}
\end{figure}

\textbf{\textit{Placement of Spc-CA and Spc-VA in Space.}} The Spc-CA and Spc-VA can be placed in any orbit or across multiple orbits. However, positioning them in MEO, between LEO and GEO, offers a strategic advantage (Figure \ref{fig:CA-VA_MEO}). This will enable broader coverage and more reliable communication links in all three orbital zones, supporting efficient certificate validation and management across space-based PKI. In order to provide a complete global connectivity between MEO and LEO, the MEO constellation of Spc-CAs/VAs must cover the entire LEO zone, with satellites positioned to ensure that at least one MEO satellite is always visible to any given LEO satellite. MEO satellites, positioned higher in space, typically cover a larger surface area than LEO satellites. This means fewer satellites are needed in MEO to cover the entire LEO, MEO, and GEO, hence, fewer Spc-CA/VA instances will be required. O3b (Other 3 Billion) constellation, operating in MEO at an altitude of ~8,000 km, aims to provide near-global broadband internet coverage using a relatively small number of 13 satellites \cite{SES-2025}. A network like the GPS constellation, which uses 24 satellites in MEO, provides full global coverage \cite{GPS-2024}. In general, a constellation may require anywhere from 10 to 25 MEO satellites to maintain constant connectivity with LEO satellites, as fewer MEO satellites are needed in this case to cover LEO orbits than to provide full Earth coverage. The precise number will depend on factors such as the desired bandwidth, latency, and the system's orbital mechanics.


\begin{figure}[ht]
    \centering
    \begin{subfigure}[b]{0.46\textwidth}
        \centering
        \includegraphics[width=\textwidth]{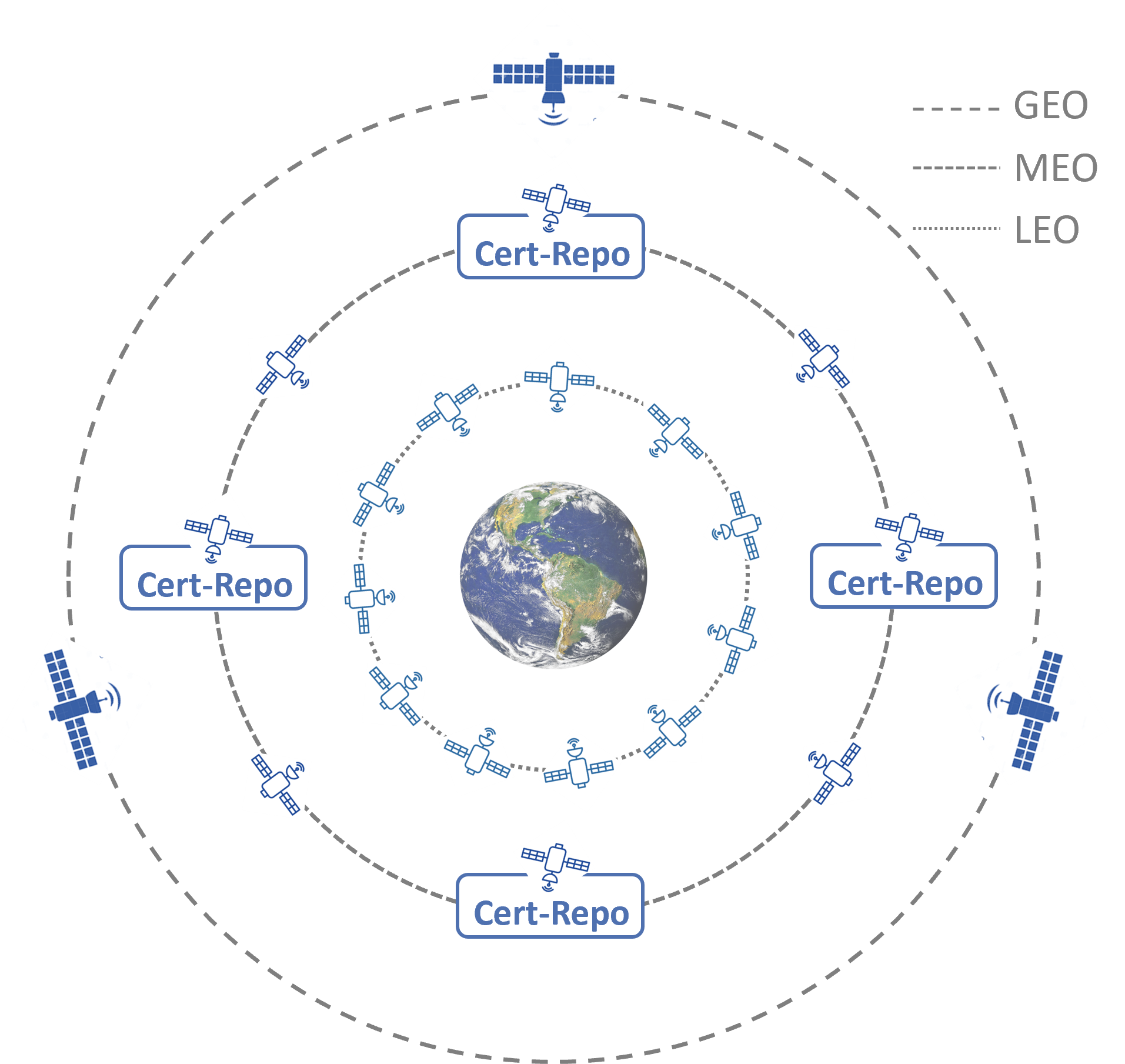}
        \caption{Repository in MEO}
        \label{fig:Cert_Repo_MEO}
    \end{subfigure}
    \hfill
    \begin{subfigure}[b]{0.45\textwidth}
        \centering
        \includegraphics[width=\textwidth]{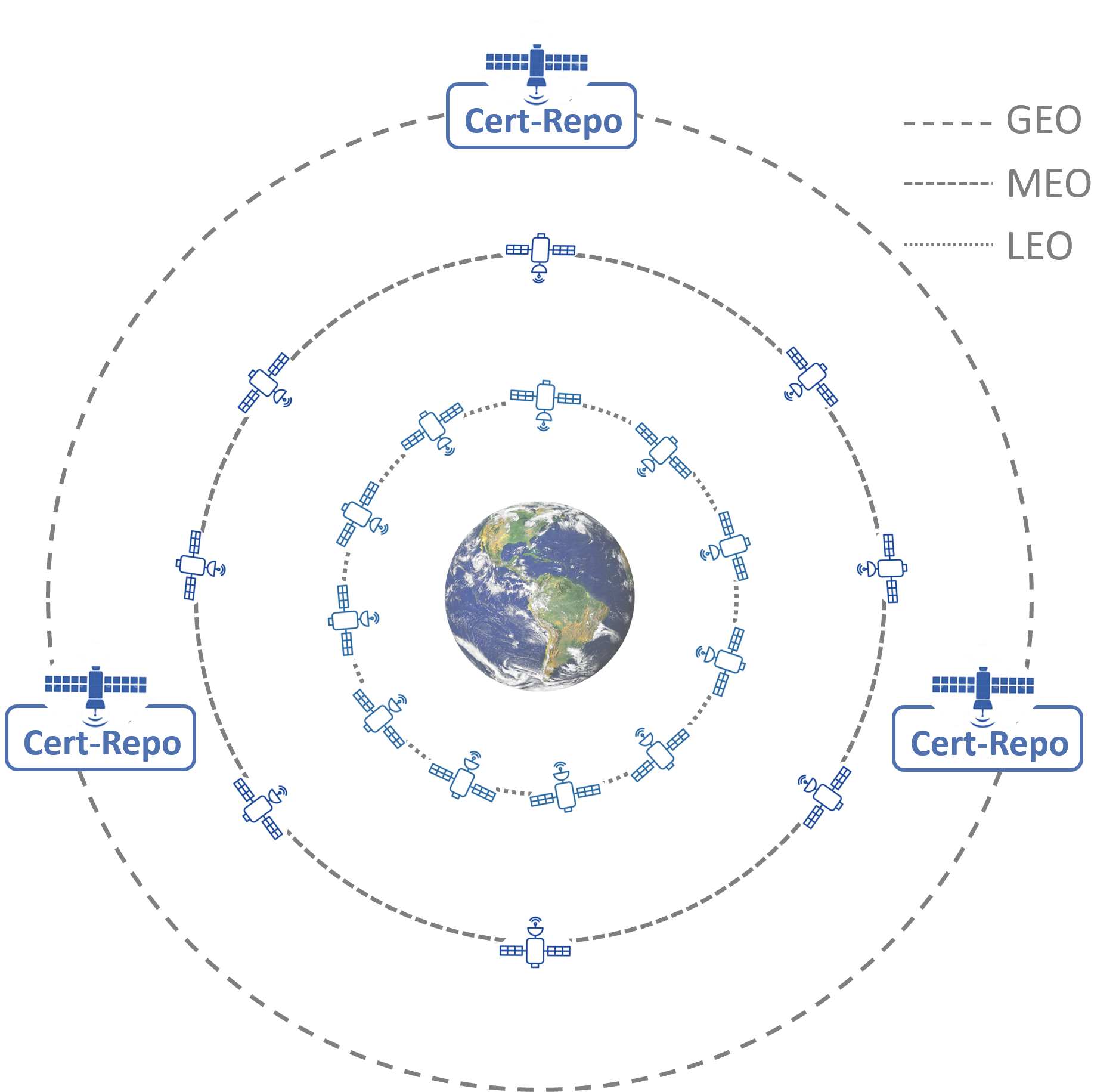}
        \caption{Repository in GEO}
        \label{fig:Cert_Repo_GEO}
    \end{subfigure}
    \caption{Placement of Certificate Repository}
    \label{fig:placement}
\end{figure}

\textbf{\textit{Placement of Certificate Repository in Space.}} The \textit{Certificate Repository} may be managed by a Spc-CA or Spc-VA, or hosted within a separate directory service, depending on the PKI architecture. In the first scenario, repository can be placed together with the Spc-CA or Spc-VA in MEO, managed by the same satellite, facilitating the deployment (Figure \ref{fig:Cert_Repo_MEO}). Certificate storage and validation will be faster with local access to the repository, particularly in SpcPKI. However, in order to facilitate real-time repository updates and reduce the number of repository instances in iPKI, it could be placed as a separate directory service in GEO (Figure \ref{fig:Cert_Repo_GEO}). Near-global coverage from GEO typically requires three satellites, spaced roughly 120° apart along the equator, to cover most of the Earth's surface \cite{CLARKE19663}. GEO satellites maintain a constant line of sight with a significant portion of the Earth's surface, enabling almost real-time updates from Grd-BCAs on the ground to the \textit{Certificate Repository} in space, addressing line-of-sight limitation of MEO satellites. They also maintain continuous connectivity with MEO satellites, ensuring that at least one GEO satellite is always visible to any given MEO satellite, facilitating Spc-VA’s access to repository. As an alternative, both the Spc-VA and the \textit{Certificate Repository} can be placed in GEO, offering global coverage along with reliable, real-time availability. 

\textbf{\textit{Certificate Format}}. The X.509 public key certificate, the most widely used format, is also the primary CCSDS-recommended credential for authentication in space \cite{CCSDS357.0-B-1}. For space-specific applications, certificates could be extended to incorporate \textit{orbital parameters} or \textit{behavioral traits} unique to each satellite. Orbital signatures, such as position, velocity, altitude, inclination, or maneuvering patterns, can serve as contextual authentication factors. Binding these traits to a satellite’s digital certificate allows to validate not only the identity of the satellite but also its expected presence in a particular orbital location, adding an extra layer of assurance against spoofing or unauthorized access.

\textbf{\textit{Certificate Validation}}. Satellites in LEO and MEO can either verify certificates themselves in SpcPKI or rely on certificate validation services provided by specialized Spc-CAs or Spc-VAs, as in both SpcPKI and iPKI. The widely known mechanisms, such as OCSP \cite{rfc6960} and OCSP stapling \cite{rfc6066}, can be utilized; however, their practical application faces significant challenges due to the unique constraints of the space environment. OCSP relies on real-time connectivity with a responder, which is an issue in space. Although OCSP stapling is better suited for such scenarios by allowing a server to ``staple'' a pre-fetched OCSP response, it can compromise revocation freshness if the status becomes outdated between updates. Additionally, OCSP provides only a certificate’s revocation status, not the full certificate or chain, which may limit effectiveness when full validation context is required. In contrast, SCVP (Server-Based Certificate Validation Protocol) \cite{rfc5055} supports transmitting not only a certificate’s status but also the full certificate or chain, enabling more comprehensive remote validation. Given these considerations, existing protocols may need to be adapted or new ones developed to ensure timely dissemination of certificates, revocation data, and policy updates to space-deployed repositories, and to support in-space certificate validation.

\textbf{\textit{Repository Updates}}. The \textit{Certificate Repository} is updated following any of these events: policy or policy mapping updates; issuance or renewal of certificates; or certificate revocation. Certificate renewal introduces new certificates, but outdated ones would not promptly harm existing operations since older, still-valid certificates remain in use for some time \cite{CCSDS357.1-O-1}, allowing periodic uploads. However, certificates are revoked more frequently than they are issued or renewed. If revocation information is delayed, systems may trust compromised certificates. In iPKI, the Grd-BCA, being the most trustworthy entity, promptly updates this information to repository after receiving it from Grd-PCAs. It may employ additional security measures and alternative faster paths to ensure the secure and timely updates. Moreover, the Grd-BCA provides the ground station facilities to establish direct communication link to the repository. Grd-BCAs will also be responsible to ensure the synchronization between \textit{Certificate Repositories}.

\textbf{\textit{Security.}} The \textit{Spc-CA}, \textit{Spc-VA}, and \textit{Grd-BCA} are \textit{trusted} services with strict security requirements. This includes secure communication channels, access control mechanisms, and encryption protocols. Secure communication protocols must be used to protect certificates and validation data during transmission, with all exchanged data authenticated using digital signatures. Cryptographic keys should be stored in secure hardware modules to prevent unauthorized access. The CCSDS has developed comprehensive standards and recommendations (e.g., \cite{CCSDS350.4-G-2}) for securing space systems, which should be followed. These include protocols specifically designed to address the unique challenges of space-based communication and the need for robust encryption and authentication. The IPSec or equivalent protocols, as recommended and adapted by CCSDS for space systems \cite{CCSDS355.0-B-2, CCSDS356.0-B-1}, can secure communication links between end user and Spc-CA, or Spc-VA, and between Spc-VA and Grd-BCA, supporting signed requests and responses. Strong security measures must also be implemented to safeguard the \textit{Certificate Repository}. Only authorized entities, such as Spc-CA, Spc-VA, and Grd-BCA, should access the repository, requiring strict access control mechanisms. A robust role-based access control system can help manage these permissions effectively. Establishing initial trust between disparate ground-based and space-based PKI systems is another key security challenge. The initial trust is established through cross-certification by the Grd-BCA in iPKI and through initial certification by the Grd-CA in SpcPKI. Later on in space, in iPKI, the end user and Spc-RP are authenticated using certificates issued by their respective space agencies, which are in trust relationships with Grd-BCA. The Spc-VA is authenticated using a certificate issued by Grd-BCA. To establish trust, space agencies must have an agreement with Grd-BCA, which allows trust in all CAs managed by the Spc-VA. Similarly, in SpcPKI, end user and Spc-RP are first authenticated to Spc-CA based on certificates issued by the Grd-CA, after which they are issued new certificates by the Spc-CA. The Spc-CA is authenticated using the certificate issued by Grd-CA. Furthermore, a Grd-PCA in iPKI may behave maliciously providing incorrect certificates or delayed revocation information; however, such behavior risks non-compliance with regulatory or contractual obligations, potentially resulting in fines, sanctions, or reputational damage. Fundamentally, trust and a strong reputation form the foundation of PKI systems.

\textbf{\textit{Scalability.}} The proposed space-based PKI systems enhance scalability by distributing trust and validation through space-based CAs and VAs, reducing reliance on a ground CAs and enabling secure operations across numerous space entities (e.g., satellites, space stations, and spacecraft). As more space users and ground entities are deployed, the PKI must scale to accommodate increasing certificates, revocation lists, and key management activities. Space-based PKI systems manage expanding trust relationships among diverse entities. Agencies can join the system through cross-certification or initial ground-based certification, while new CAs or domains can be integrated by publishing certificates, updated cross-certifications, and policies to \textit{Certificate Repository}. To maintain efficient operation in this expanding environment, new trust relationships must be validated in real time, allowing new certificate users and Space-RPs to seamlessly access PKI services. Scalability is further supported by interoperability, achieved through common standards such as X.509 certificates and protocols like OCSP and CRL. This approach ensures diverse systems from multiple countries and organizations can interoperate without major redesigns, while reducing dependence on GEO satellites and ground infrastructure that could otherwise become bottleneck. As space networks grow, managing certificate lifecycles becomes more complex, especially for revocation. Space-based PKI addresses this by hosting the \textit{Certificate Repository} in space and automating issuance, renewal, and revocation. Ultimately, scalability ensures PKI systems can securely support growing numbers of users.

\textbf{\textit{Cost Considerations and Business Model.}} While deploying a PKI enhances security and trust within space communications, it also involves associated costs. The main cost components include hardware and software, implementation (design, installation, and configuration), operations, and maintenance. In iPKI, only VAs are deployed in space reducing hardware requirements, however, it introduces added complexity in maintaining operability across different systems. By contrast, SpcPKI is conceptually more straightforward, since it unifies all space entities under a single PKI system. However, it requires more extensive infrastructure in space to realize its full potential. The cost of setting up and operating the Grd-CAs in iPKI will be born by the space agencies that rely upon those certificates as is done in the current state. For the trusted services provided by space-based PKI systems, the end user and Spc-RP (i.e., respective space agencies) will pay according to the business model agreement and service requirements (e.g., transaction based, volume based, or fixed) agreed at the registration time. For Spc-RPs receiving certificates from various sources (e.g., alerts from collision avoidance and space traffic management systems), performing certificate validation locally or outsourcing it to a Spc-CA or Spc-VA is more cost-effective than sending each request to the ground. Without space-based PKI systems, each certificate validation request would require the Spc-RP to either wait for a ground station pass or rely on paid data relay services, such as those provided by NASA's relay system \cite{TDRS-2023}. In some cases, they may also need to pay for the certificate validation services at ground. By subscribing to space-based PKI services, the RP obtains timely access to accurate and up-to-date validation information, which is essential for making informed and trustworthy decisions. The value provided to RPs is compelling enough to justify their willingness to pay for it. The existing data relay systems in GEO can be (re-)used to offer repository in space as a paid service if \textit{Certificate Repository} is to be deployed in GEO. On the other hand, there are satellite systems operational in MEO, e.g., the GPS satellite system \cite{GPS-2024}, which could provide additional cost effective PKI services. While building a space-based PKI mechanism may require significant \textit{upfront investment}, the long-term benefits including lower operational costs, improved efficiency, and streamlined collaboration, far outweigh the initial expenses. Moreover, it aligns with international cybersecurity objectives and provides the foundation for secure, interoperable, and scalable next-generation space missions.

\subsection{Comparing Latency in Existing and Proposed PKI Approaches}

Certificate validation in a PKI system is mainly evaluated by \textit{latency} and \textit{efficiency}, which directly impact overall performance. As mentioned earlier, technical implementation and evaluation of proposed PKI systems are not the focus of this paper. However, comparing certificate validation results, especially the \textit{estimated communication latency} below, provides insights into the performance of existing approaches and the proposed PKI systems. \textit{Latency} refers to the communication (signal propagation) delay from the initiation of a validation request to the delivery of the result, excluding the processing time. For simplicity, factors that may affect latency, such as communication technologies used, digital processing delays (e.g., packet processing, data encoding/decoding, certificate verification, and path validation), signal attenuation, multi-hop communication, and delays within the terrestrial part of the network, are not considered. \textit{Latency} is estimated based on the space communication delay between the two endpoints, computed as:

\begin{center}
\textit{One-way latency} = \textit{Distance / Speed of Light}
\end{center}

Where \textit{Distance} is the separation between the two communication points (e.g., between Earth and LEO) and \textit{Speed of Light} is approximately 299,792 km/s in a vacuum.

\subsubsection{Delayed Validation}
\label{validtion_Delayed}

As discussed in Section \ref{AuthPath}, LEO and MEO satellites must wait until their next pass over an approved ground station to perform certificate validation. 

\begin{itemize}
     
    \item \textit{LEO-to-ground communication latency.} For satellites in LEO, the orbital period is approximately 90 to 120 minutes, depending on their altitude. The travel time for communication signals to and from the satellite to Earth (round-trip latency) is $\sim$1 ms at 160 km altitude to $\sim$14 ms at 2,000 km altitude.
        
    \item \textit{MEO-to-ground communication latency.} For MEO, the orbital period is approximately 5 to 12 hours (e.g., 12 hours for GPS satellites at 20,200 km altitude). The round-trip latency ranges from $\sim$56 ms at 2,000 km to $\sim$134 ms at 20,200 km altitude.

\end{itemize}


\textit{\textbf{Remarks.}} The communication latency for direct LEO/MEO-to-ground links is low, with orbital periods of $\sim$120 minutes for LEO and $\sim$12 hours for MEO. However, due to the Earth's rotation, they may not visit the same ground station every time. Considering the satellites' orbital inclinations, along with the combined effects of their speeds and Earth’s rotation, they may take up to 24 hours to pass over the same ground station, resulting in a wait time of up to 24 hours. Reducing this wait time would require multiple ground stations strategically positioned worldwide. Moreover, LEO satellites ($\sim$300-500 km) are visible for about 5-10 minutes per pass, while MEO satellites ($\sim$2,000-10,000 km) are visible for roughly 20 minutes to 2 hours. After this time, they will have to connect to the next ground station. Continuous connectivity would necessitate a large network of ground stations to maintain line-of-sight for real-time certificate verification, making it both costly and politically challenging.

\subsubsection{Validation via Data Relay Satellites}
\label{validtion_Relay}

As described in Section~\ref{AuthPath}, a LEO satellite sends the certificate to a GEO satellite, which relays it to the ground for verification and returns the result back to the LEO satellite.

\begin{itemize}
 
    \item \textit{LEO-to-GEO-to-ground communication latency.} For a LEO at 1,000 km and GEO at 35,780 km, the estimated LEO-GEO round-trip latency is $\sim$232 ms, assuming the LEO satellite is directly beneath the GEO satellite. For GEO, the GEO-to-ground round-trip latency will be $\sim$260 ms for a ground station located at a mid-latitude location \cite{Book-2008}. Adding the  two latencies results in $\sim$492 ms, the estimated certificate verification delay, excluding all other processing required to verify a certificate.
 
\end{itemize}

\textit{\textbf{Remarks.}} Verification via GEO relay satellites adds LEO-GEO-ground round-trip latency, which is further affected by additional factors. Due to relative motion, the distance between a LEO and a GEO satellite begins to increase as the LEO satellite moves away from a position directly beneath the GEO satellite. This difference is particularly significant when the LEO satellite is over mid-latitude regions or in a polar orbit above the polar regions. In such cases, the LEO satellite may lose line of sight to the GEO satellite, reducing availability of validation service. Moreover, current GEO relay systems typically serve a limited number of satellites at a time, e.g., NASA’s TDRS can serve up to 20 spacecraft simultaneously \cite{TDRS-capacity}. Their main purpose is to quickly relay LEO/MEO satellite data to the ground, critical for applications like remote sensing, weather forecasting, and surveillance. They achieve this via time-sharing, where the relay system allocates time slots to LEO/MEO satellites instead of providing continuous dedicated channels. On the uplink, the ground station can only transmit when the satellite permits it; otherwise, collisions may occur. Hence, the LEO-to-GEO-to-ground round-trip latency further increases due to queuing on the upstream and downstream. Furthermore, longer GEO-ground signal travel times also increase the risk of data loss \cite{Book-2008}. Snow, rain, clouds, and ice attenuate radio signals, further increasing data loss. As a result, each satellite's transmission rate is constrained, posing challenges for real-time certificate validation.


\subsubsection{Validation in iPKI and SpcPKI}

\textit{Space-to-space communication latency} depends on the deployment of \textit{Spc-CA/Spc-VA} and \textit{Certificate Repository}. Two potential deployment scenarios are considered to illustrate practical choices for maximizing coverage and minimizing latency.

\medskip

\begin{figure}[ht]
\centering
\begin{subfigure}[b]{0.45\textwidth}
    \centering
    \includegraphics[width=\textwidth]{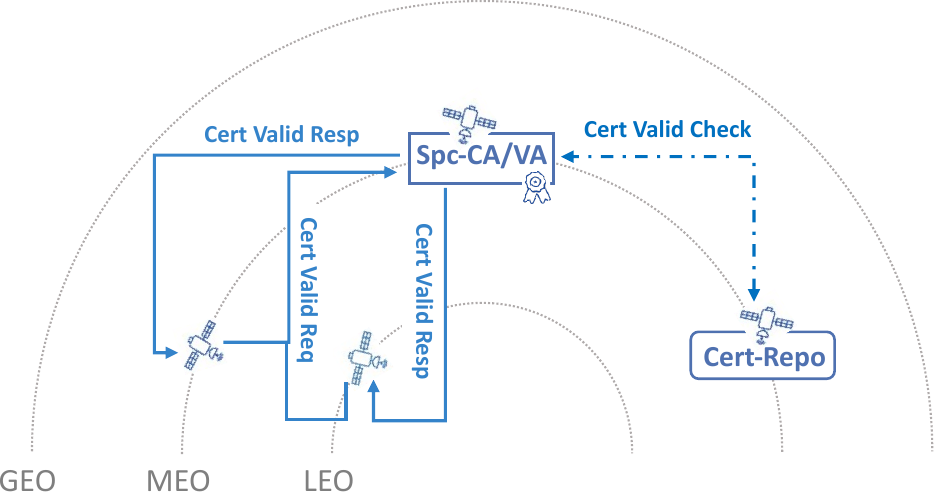} 
    \caption{Repository in MEO}
    \label{fig:Cert_Validation_MEO}
\end{subfigure}
\hspace{1cm}
\begin{subfigure}[b]{0.45\textwidth}
    \centering
    \includegraphics[width=\textwidth]{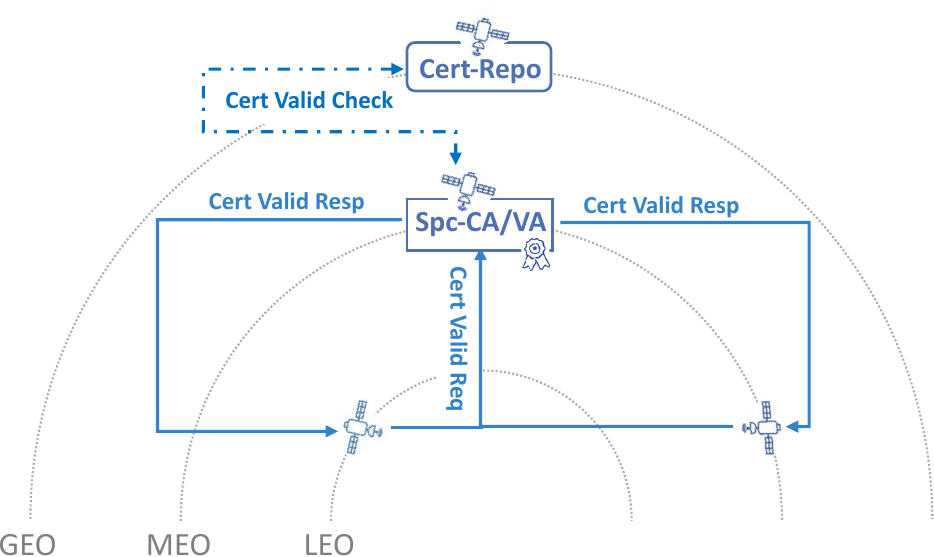}
    \caption{Repository in GEO}
    \label{fig:Cert_Validation_GEO}
\end{subfigure}

\caption{Certificate Validation With Certificate Repository in MEO and GEO}
\label{fig:Cert_Validation}
\end{figure}

\noindent\underline{\textbf{\textit{Case 1.}} \textit{Spc-CA/Spc-VA and Certificate Repository in MEO}}. A certification validation request from a LEO/MEO satellite is processed by the \textit{Spc-CA/Spc-VA} located at an altitude of 10,000 km in MEO with access to the \textit{Certificate Repository} also deployed in MEO (Figure \ref{fig:Cert_Validation_MEO}). This altitude allows visibility of many LEO satellites while enabling efficient communication with both LEO and higher MEO satellites. Other altitudes may be chosen based on coverage, latency, and network design.

\begin{itemize}
    \item \textit{LEO-to-MEO communication latency.} For the Spc-RP satellite at 1,000 km in LEO and $\sim$9,000 km away from the \textit{Spc-CA/Spc-VA}, the \textit{round-trip latency} between Spc-RP and Spc-CA/Spc-VA is (2x30 ms) $\sim$60 ms only.

    \item \textit{MEO-to-MEO communication latency.} For the Spc-RP satellite at 5,000 km in MEO and $\sim$5,000 km away from the \textit{Spc-CA/Spc-VA}, the \textit{round-trip latency} between Spc-RP and Spc-CA/Spc-VA is (2x16.7 ms) $\sim$33 ms only.
      
\end{itemize}

\noindent\underline{\textbf{\textit{Case 2.}} \textit{Spc-VA in MEO and Certificate Repository in GEO}}. This scenario can be deployed for iPKI, which involves ground-space communication for certificate updates. A certification validation request from a LEO/MEO satellite is processed by the \textit{Spc-VA} located at an altitude of 10,000 km in MEO with access to the \textit{Certificate Repository} located at an altitude of 35,780 km in GEO (Figure \ref{fig:Cert_Validation_GEO}). This setup will add the MEO-to-GEO communication latency to previous case that will be (2x85.99 ms) $\sim$172 ms.

\begin{itemize}
    \item \textit{LEO-to-MEO-to-GEO communication latency.} If the Spc-RP satellite is positioned at an altitude of 1,000 km in LEO, the round-trip communication latency from LEO to MEO to GEO will be (60+172 ms) $\sim$232 ms.

    \item \textit{MEO-to-MEO-to-GEO communication latency.} Similarly, if the Spc-RP satellite is positioned at 5,000 km in MEO, the round-trip communication latency from MEO to MEO to GEO will be (33+172 ms) $\sim$205 ms.
        
\end{itemize}



\textit{\textbf{Remarks.}} The validation time can be very short when a certificate is verified locally by the Spc-RP within SpcPKI, enabling near real-time verification. For third-party validation by CAs or VAs, the validation time depends on factors such as orbits, altitudes, distances, and other deployment choices. In both cases above, the maximum latency occurs when the Spc-VA is in MEO and the certificate repository in GEO ($\sim$232 ms), which is still lower than relay-based validation latency ($\sim$492 ms).

\section{Related Work}
\label{related}




\subsection{Symmetric Key-Based Authentication}

The ESA defines a \textit{shared-key-based authentication} mechanism for verifying telecommands from ground stations to spacecraft \cite{ESAPSS-04-107,ESAPSS-04-151}. The telecommand data is first hashed using the secret key shared between the two parties. It is then processed through a one-way function to generate an ``authentication signature''. Upon receiving the telecommand, spacecraft repeats the same process to ensure legitimacy of sender and the contents. For replay attacks, both parties maintain a counter, which is incremented after each successful authentication. The ESA's Copernicus telecommand authentication system \cite{CCSDS350.0-G-3}, currently deployed in the Copernicus Sentinel fleet, further includes \textit{functionality for uploading new authentication keys to the spacecraft}. In line with the CCSDS Symmetric Key Management Recommended Standard \cite{CCSDS354.0-M-1}, the system uses a two-tier key hierarchy with master and session keys: session keys authenticate telecommands and are periodically updated via Over-The-Air Rekeying (OTAR), while master keys protect session keys during rekeying and support other critical functions. All key generation and management is handled on the ground. 

\medskip

\underline{\textit{Remarks.}} Symmetric key-based authentication works well in closed environments such as ESA, NASA, or the ISS, where only a limited number of trusted entities are involved. However, in more dynamic space communication environments with many diverse stakeholders, such mechanisms become less effective. The complex relationships between stakeholders call for more flexible and scalable PKI-based solutions to ensure both security and interoperability.

\subsection{PKI-Based Authentication}



\subsubsection{PKI Interoperability} 

\begin{figure}[ht]
    \centering
    \includegraphics[width=0.8\textwidth]{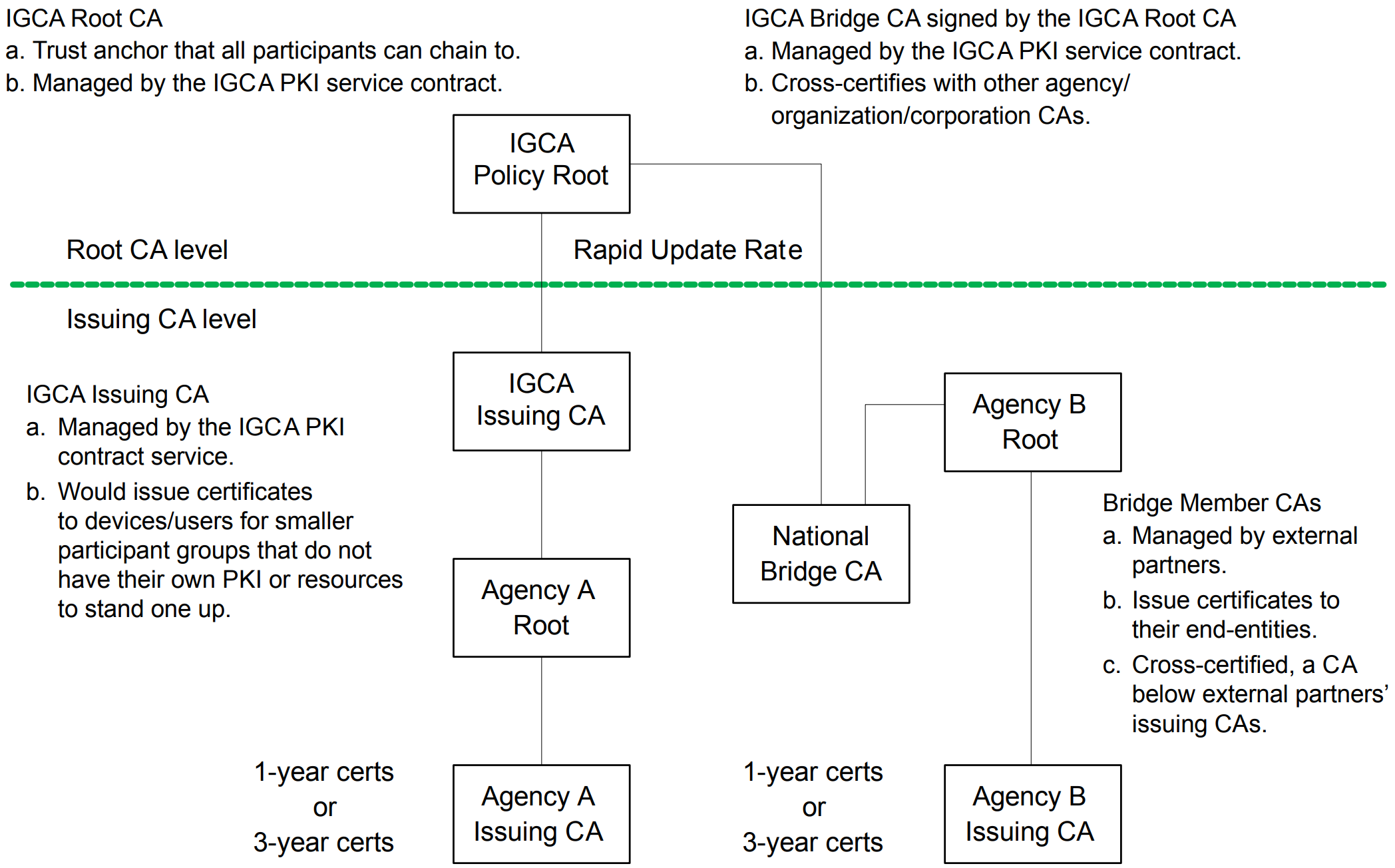}
    \caption{IGCA Bridge CA by CCSDS \cite{CCSDS357.1-O-1}}
    \label{fig:CCSDS_BCA}
\end{figure}

CCSDS’s work on PKI interoperability for space missions is among the most relevant contributions in this area. CCSDS highlights that secure space operations rely on PKI-based interoperability to enable cross-support between agencies and their partners. To facilitate this, CCSDS proposed the Intergovernmental Certification Authority (IGCA) in \cite{CCSDS357.1-O-1}, defining how the IGCA can act as a Bridge CA between member space agencies, enabling CAs to issue digitally signed certificates that can be used to secure access and communications paths. The IGCA functions as a federated PKI framework. At its core, the IGCA Root CA serves as the ultimate trust anchor that all participants can chain to. The IGCA Bridge CA, signed by the IGCA Root CA, establishes interoperability by cross-certifying with agency PKIs, allowing their certificates to be mutually trusted. In addition, an IGCA Issuing CA is included to provide end-entity certificates for smaller agencies, projects, or participant groups that lack their own PKI. Together, these components ensure certificate validation, dissemination, and secure interoperability for space missions (see Figure~\ref{fig:CCSDS_BCA}). The IGCA acts as the central authority for a CCSDS-wide PKI, enabling universal international mission security without overriding individual nations’ certificate or key management controls. The IGCA has been proposed to support the Lunar Communications Architecture by enabling certificate validation, certificate dissemination, and other security functions \cite{CCSDS357.1-O-1}.

\medskip

\underline{\textit{Remarks.}} While the IGCA provides a ground-based framework for certificate issuance and dissemination across multiple space agencies, it does not define mechanisms for in-space verification of these certificates. As a result, the authentication of satellites or space systems from different agencies during actual in-orbit operations remains unaddressed.

\subsubsection{Key Management} 

\begin{figure}[h]
    \centering
    \includegraphics[width=0.60\textwidth]{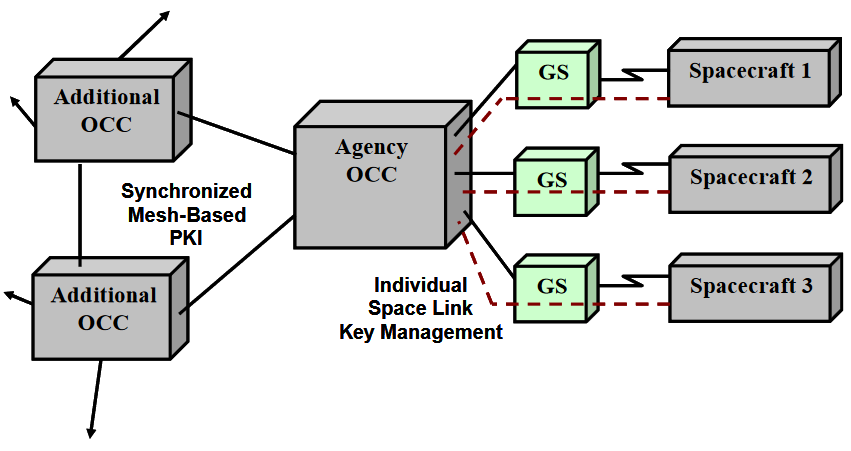}
    \caption{Constellation with Individual Key Management \cite{CCSDS350.6-G-1}}
    \label{fig:ccsdsmeshpki}
\end{figure}

The CCSDS Space Missions Key Management Concept \cite{CCSDS350.6-G-1} defines spacecraft \textit{key management} by combining ground-based PKI with symmetric keys on space links. In a typical mission, key management typically involves the spacecraft and a ground Operations Control Center (OCC). Cryptographic keys must be securely distributed and synchronized between the OCC and each spacecraft, with all key generation, whether symmetric or asymmetric, performed by the OCC or a co-located authority. In constellations, a single OCC may serve as a central authority. In constellations managed by multiple OCCs, each OCC operates as a node within a mesh-based PKI system, enabling secure inter-OCC and space-ground communication (see Figure~\ref{fig:ccsdsmeshpki}). 

\medskip

\underline{\textit{Remarks.}} The report addresses only key management and does not cover certificate validation or management in single- or multi-OCC setups.

\subsubsection{Certificate Validation}

\textit{KeySpace} \cite{keyspace} proposes a PKI evaluation framework for interplanetary networks and studies how conventional mechanisms such as OCSP and CRLs can operate under high-latency, disrupted connectivity. It demonstrates that distributing authority functions and leveraging relay nodes can improve certificate validation and connection establishment performance in delay- and disruption-tolerant space environments. While KeySpace complements this direction by showing the adaptability of traditional PKI mechanisms under space-like constraints, it primarily focuses on evaluation and adaptation rather than defining a dedicated space-native PKI architecture, highlighting the need for purpose-built space-based PKI frameworks.

A practical PKI design tailored for space networks is presented in \cite{Revoc}, with a focus on \textit{improving efficiency in certificate management} and secure communication. It proposes a decentralized PKI architecture that uses peer-to-peer epidemic dissemination to handle high latency and disruption in space communication. By replacing traditional, connection-dependent CRLs and OCSP with efficient, cryptographically verifiable revocation data, the design supports secure multi-authority operations in complex satellite constellations while reducing network overhead. The work shows how existing PKI mechanisms can be adapted to operate more efficiently in space environments. However, the approach still follows a largely ground-centric PKI model, where certificate issuance and validation depend on terrestrial infrastructure. As a result, it does not fully address the limitation of delayed validation in highly dynamic and multi-operator constellations, where in-orbit trust establishment and autonomous validation are increasingly required.

\begin{figure}[h]
    \centering
    \includegraphics[width=0.60\textwidth]{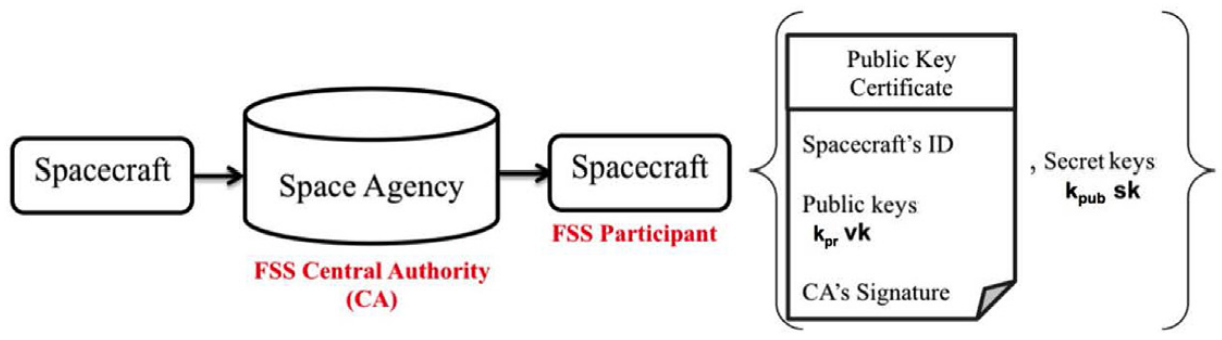}
    \caption{Federated Satellite System PKI \cite{VONMAURICH2018}}
    \label{fig:fsspki}
\end{figure}

A PKI-based \textit{identity verification protocol} for federated satellite systems (FSSs) was proposed in \cite{VONMAURICH2018} to ensure data authentication, integrity, and confidentiality in satellite-to-satellite communication. Using this system, each spacecraft within a FSS can verify the identity of other member spacecraft by validating their public key certificates. To join the network, a spacecraft must first registers with the CA of FSS. During the registration, the CA confirms the spacecraft's identity by verifying the operator's credentials. Once validated, the CA issues a signing and verification key pair along with a public key certificate (see Figure~\ref{fig:fsspki}). This certificate contains the spacecraft’s identity, public keys. Spacecraft authenticate each other by validating certificates with the CA's public key. For verification, the relying party must also be part of the federated system and possess a trusted copy of the CA's public key. Since both the certificate holder and verifier must operate within the same FSS PKI, certificates from external PKI systems cannot be validated in this setup. 

\smallskip

For \textit{data authentication} in Satellite-Based Augmentation Systems (SBAS), a PKI-based key management framework is proposed in \cite{Neish2019}. SBAS improves the accuracy and reliability of GNSS positioning data, with information transmitted in a unidirectional stream from satellite to user. This one-way communication makes the authentication requirement one-directional. For this purpose, the proposed framework relies on a two-level key hierarchy. The \textit{Level 1} keys, embedded in the receiving devices at the time of manufacturing, act as the root of trust. The \textit{Level 2} public keys, which are used to authenticate positioning data messages, are broadcast over the air signed by \textit{Level 1} keys. Pre-installed \textit{Level} 1 keys eliminate the need for certificate validation in the one-way satellite-to-user stream.

\medskip

\underline{\textit{Remarks.}} These examples represent PKI applications in space, but existing systems are typically single-network and neglect certificate management in dynamic space environments.

\section{Open Challenges and Future Directions}
\label{future-work}

This work aims to establish the need for and feasibility of space-based public key systems, rather than providing a complete implementation or security evaluation. The main objective is to define the design space and identify the constraints that directly affect certificate-based trust in space. These constraints must be clearly understood before concrete protocols and implementations can be developed, making this a necessary first step toward secure and scalable cryptographic solutions for future space networks.

A key observation of this study is the limitation of ground-dependent certificate validation. Relying on ground-based validation introduces latency, reduces system responsiveness, and creates operational dependencies that are not suitable for time-critical satellite operations. Enabling certificate validation within the space segment allows satellites to authenticate data and commands locally, which is essential for real-time inter-satellite communication and cross-operator coordination. At the same time, moving trust services into space introduces new challenges. Traditional PKI systems assume persistent connectivity to certificate authorities and validation services. This assumption does not hold in space due to intermittent links, line-of-sight constraints, and long delays. As a result, these systems cannot be directly migrated to space as-is and require significant adaptation. 

The latency analysis presented is intended as a preliminary and illustrative assessment of the communication implications of ground-dependent versus space-based trust models. While it does not capture the full complexity of satellite network behavior, it highlights the potential benefits of in-space validation. A more comprehensive evaluation would need to consider constellation dynamics, inter-satellite link delays, routing strategies, network congestion, and revocation propagation across distributed nodes. Such studies typically require detailed simulation frameworks or system-level experimentation and are left for our future work. Similarly, the proposed framework does not provide concrete protocol instantiations or a formal security analysis. These aspects depend on the architectural choices and system requirements established at this stage. The goal of this work is to motivate and structure the design space for space-based PKI systems, rather than to finalize specific implementations. Once the architectural framework is defined, future efforts will focus on protocol design, optimization of cryptographic parameters, and evaluation under realistic operational conditions.

Space-based PKI systems must operate under dynamic network topologies, constrained onboard resources, and heterogeneous, multi-authority environments. These factors directly affect how certificates are issued, stored, distributed, validated, and revoked. Conventional mechanisms are not efficient under these conditions and require adaptation. Identifying these limitations helps clarify the need for new or adapted approaches to decentralized validation, efficient certificate dissemination, and robust trust management in distributed and intermittently connected networks. The proposed framework highlights several open research challenges. The realization of the proposed deployment schemes requires the development of concrete protocols and system mechanisms tailored to space environments. In particular, the proposed schemes lead to the following technical directions:

\begin{itemize}
    \item \textit{\textbf{In-orbit validation protocols.}} Current on-ground deployments rely on OCSP for real-time certificate status queries. This model is not suitable for intermittent connectivity. A practical direction is to design a delay-tolerant OCSP variant, e.g., \cite{ADOPT_OCSP}, where validation authorities pre-compute and sign status responses with short validity intervals (e.g., minutes to hours) and distribute them across satellites. These responses can be cached and verified locally. In addition, push-based dissemination over inter-satellite links can replace query-based validation, using periodic broadcast or gossip-based protocols like \cite{Revoc}. To maintain consistency, a state synchronization protocol among VAs is required, potentially using versioned status updates and conflict resolution based on timestamps or signed epochs.
    
    \item \textit{\textbf{Revocation mechanisms under intermittent connectivity.}} Standard CRLs are large and require frequent updates. A more suitable approach is to use delta-CRLs \cite{delta_CRLs} combined with Merkle tree commitments \cite{Merkle_Tree}, where satellites verify revocation status using compact proofs. Bloom filter-based revocation summaries can provide fast membership checks with bounded false positives \cite{CRLite}. Revocation updates can be propagated using store-carry-forward (DTN-style) \cite{DTN} dissemination, ensuring eventual consistency. Protocols should define explicit staleness thresholds (e.g., maximum acceptable revocation delay) and fallback policies when fresh revocation data is unavailable.
    
    \item \textit{\textbf{Certificate distribution and repository design:}} Instead of centralized repositories, certificates can be stored in a distributed hash table (DHT)-like structure \cite{DHT} mapped onto satellite constellations, where nodes are responsible for subsets of certificates. Replication can follow orbital planes to ensure availability across visibility windows. Opportunistic caching along communication paths can reduce lookup latency. Certificate retrieval can be integrated with routing using name-based \cite{NDN} or content-centric \cite{CCN} approaches (e.g., ICN/NDN-style retrieval \cite{DistributedOCSP,NDNCert}), allowing certificates to be fetched from any reachable node that holds a valid copy.
    
    \item \textit{\textbf{Distributed certificate authority protocols (fully in-space scheme).}} In the fully in-space model, certificate issuance must be decentralized. This can be achieved using threshold signature schemes \cite{ThresholdDSS} (e.g., BLS threshold signatures or threshold ECDSA), where a subset of satellites collaboratively signs certificates. Distributed key generation protocols can be used to establish CA keys without a trusted dealer. Certificate issuance can follow a quorum-based protocol (e.g., \cite{COCA, QM_Proto}), where a minimum number of satellites approve and co-sign a request. To handle disrupted connectivity, asynchronous consensus or partially synchronous protocols (e.g., PBFT variants adapted for delay \cite{PBFT, PBFT2, BFT}) can be used with delayed commit semantics.
    
    \item \textit{\textbf{Trust synchronization and multi-authority management.}} In multi-operator environments, trust anchors must be synchronized efficiently. This can be achieved using incremental trust updates with signed trust bundles, where only changes are propagated. e.g., \cite{rfc5934}. Certificate path validation can be optimized using pre-validated trust paths or short-lived cross-certificates to reduce verification overhead. For interoperability, bridge CA models or cross-certification protocols can be adapted adding additional mechanisms to handle delayed updates and partial trust views.
    
    \item \textit{\textbf{Cryptographic optimization for space platforms.}} Cryptographic choices must balance security and efficiency. For signatures, Ed25519 or ECDSA (P-256) offer efficient verification and compact sizes for current deployments. For aggregation and distributed signing, BLS signatures are attractive due to their support for aggregation, reducing communication overhead. For post-quantum readiness, schemes such as CRYSTALS-Dilithium \cite{CRYSTALS-Dilithium} can be evaluated, though their larger sizes require careful optimization. Hybrid certificates combining classical and post-quantum signatures can provide a transition path. Certificate formats can also be optimized using compressed encodings (e.g., CBOR-based certificates \cite{CBOR}).
 
\end{itemize}

These approaches provide directions for enabling space-based public key systems, but they are not exhaustive and do not represent finalized solutions. They highlight how trust services can be moved into the space segment and how existing mechanisms may be adapted or redesigned under space constraints. Some techniques draw from existing networking and cryptographic methods, but their effectiveness under space-specific constraints remains to be validated. Certain approaches may require refinement, combination with other methods, or replacement based on detailed evaluation and system-level experimentation. The realization of space-based public key systems requires coordinated advances in protocol design, distributed systems, and cryptographic engineering. This work establishes the architectural foundation and highlights key technical directions. The proposed PKI architecture remains compatible with current public-key cryptography while being structurally adaptable to future post-quantum cryptographic mechanisms, such as lattice-based schemes (e.g., Kyber and Dilithium), as standards mature.

Future work will focus on developing concrete, space-aware protocols for certificate issuance, validation, distribution, and revocation. This includes refining the proposed techniques, combining them where needed, and designing mechanisms that better fit disruption-tolerant environments. In addition, implementing the proposed schemes on realistic simulation platforms will be important to assess practical feasibility. Further evaluation under large-scale constellation settings will provide deeper insight into latency, scalability, and system behavior. Formal security analysis and protocol validation will help assess resilience against advanced threats. Moreover, integrating these designs with existing standards and cross-agency frameworks will be essential to support interoperability and collaborative space missions. 

As an immediate next step, we will carry out a detailed system-level evaluation and security analysis. This will involve integrating PKI operations into satellite network simulators or DTN-based frameworks to capture mobility, intermittent links, and routing behavior. We will analyze key metrics such as validation latency, revocation delay, bandwidth overhead, and computational cost. In parallel, we will model adversarial scenarios, including compromised satellites, delayed revocation, and inconsistent trust states, to evaluate the robustness of the proposed approaches.


\section{Conclusion}
\label{conclusion} 

This paper presents, to best of our knowledge, the first conceptual designs for space-based PKI to secure satellite operations and address limitations of ground-dependent approaches. Existing PKI systems rely heavily on ground-based certificate validation, causing delays and security risks. Space-based PKI moves certificate management and validation into space, \textit{reducing reliance on ground stations} and  \textit{lowering communication latency}. Two PKI schemes were proposed: an integrated PKI that incrementally extends existing infrastructure, and a fully in-space PKI that offers greater autonomy and scalability. Both schemes \textit{enable interoperability and enhanced collaboration among missions and agencies}, providing a foundation for resilient and trusted operations across the space ecosystem. Baseline latency analysis highlights the potential for improved real-time decision-making and more efficient satellite operations.

\bibliographystyle{IEEEtran}
\bibliography{Space-Based-PKI}

@article{GOLKAR2015230,
title = {The Federated Satellite Systems paradigm: Concept and business case evaluation},
journal = {Acta Astronautica},
volume = {111},
pages = {230-248},
month = {Jun.},
year = {2015},
issn = {0094-5765},
doi = {https://doi.org/10.1016/j.actaastro.2015.02.009},
author = {Alessandro Golkar and Ignasi {Lluch i Cruz}},
}

@incollection{CLARKE19663,
title = {Extra-Terrestrial Relays: Can Rocket Stations Give World-wide Radio Coverage?},
editor = {Richard B. Marsten},
series = {Progress in Astronautics and Rocketry},
publisher = {Elsevier},
volume = {19},
pages = {3-6},
year = {1966},
booktitle = {Communication Satellite Systems Technology},
issn = {0079-6050},
doi = {https://doi.org/10.1016/B978-1-4832-2716-0.50006-2},
author = {ARTHUR C. CLARKE},
}

@inbook{STM2019,
author = {David D. Murakami and Sreeja Nag and Miles Lifson and Parimal H. Kopardekar},
title = {{Space Traffic Management with a NASA UAS Traffic Management (UTM) Inspired Architecture}},
booktitle = {AIAA Scitech 2019 Forum},
chapter = {},
pages = {},
publisher = {American Institute of Aeronautics and Astronautics (AIAA)},
year = {2019},
doi = {10.2514/6.2019-2004},
}

@Conference{intercomm2001,
  author       = {Kul Bhasin and Jeffrey Hayden},
  title        = {Inter-spacecraft communication architectures and technologies for coordinated spacecraft missions},
  booktitle    = {AIAA Space 2001 Conference and Exposition},
  year         = {2001},
  month        = {August},
  address      = {Albuquerque, NM, USA},
  doi          = {https://doi.org/10.2514/6.2001-4709},
  url          = {https://arc.aiaa.org/doi/pdf/10.2514/6.2001-4709}
}

@techreport{NASA2021,
type = {Report},
author = {{NASA Office of Inspector General}},
title = {{NASA’S Efforts To Mitigate The Risks Posed By Orbital Debris}},
month = {January},
year = {2021},
institution = {NASA},
address = {Washington, DC, USA},
url = {https://oig.nasa.gov/wp-content/uploads/2024/02/IG-21-011.pdf},
}

@techreport{CCSDS350.6-G-1,
    type = {Standard},
    title       = {{CCSDS Report Concerning Space Data System Standards: SPACE MISSIONS KEY MANAGEMENT CONCEPT}},
    month       = {Nov.},
    year        = {2011},
    number      = {CCSDS 350.6-G-1},
    institution = {Consultative Committee for Space Data Systems (CCSDS)},
    address     = {Washington, D.C., USA},
    url         = {https://ccsds.org/Pubs/350x6g1.pdf},
    note        = {{Informational Report, Issue 1}}
}

@techreport{CCSDS350.0-G-3,
    type = {Standard},
    title       = {{CCSDS Report Concerning Space Data System Standards: THE APPLICATION OF SECURITY TO CCSDS PROTOCOLS}},
    month       = {Mar.},
    year        = {2019},
    number      = {CCSDS 350.0-G-3},
    institution = {Consultative Committee for Space Data Systems (CCSDS)},
    address     = {Washington, D.C., USA},
    url         = {https://ccsds.org/Pubs/350x0g3.pdf},
    note        = {{Informational Report, Issue 3}}
}

@techreport{CCSDS350.4-G-2,
    type = {Standard},
    title       = {{CCSDS Report Concerning Space Data System Standards: CCSDS GUIDE FOR SECURE SYSTEM INTERCONNECTION}},
    month       = {Apr.},
    year        = {2019},
    number      = {CCSDS 350.4-G-2},
    institution = {Consultative Committee for Space Data Systems (CCSDS)},
    address     = {Washington, D.C., USA},
    url         = {https://ccsds.org/Pubs/350x4g2.pdf},
    note        = {{Informational Report, Issue 2}}
}

@techreport{CCSDS351.0-M-1,
    type = {Standard},
    title       = {{CCSDS Recommendation for Space Data System Practices: SECURITY ARCHITECTURE FOR SPACE DATA SYSTEMS}},
    month       = {Nov.},
    year        = {2012},
    number      = {CCSDS 351.0-M-1},
    institution = {Consultative Committee for Space Data Systems (CCSDS)},
    address     = {Washington, D.C., USA},
    url         = {https://ccsds.org/Pubs/351x0m1.pdf},
    note        = {{Recommended Practice, Issue 1}}
}

@techreport{CCSDS352.0-B-2,
    type = {Standard},
    title       = {{CCSDS Recommendation for Space Data System Practices: CCSDS CRYPTOGRAPHIC ALGORITHMS}},
    month       = {Aug.},
    year        = {2019},
    number      = {CCSDS 352.0-B-2},
    institution = {Consultative Committee for Space Data Systems (CCSDS)},
    address     = {Washington, D.C., USA},
    url         = {https://ccsds.org/Pubs/352x0b2.pdf},
    note        = {{Recommended Standard, Issue 2}}
}

@techreport{CCSDS354.0-M-1,
    type = {Standard},
    title       = {{CCSDS Recommendation for Space Data System Practices: SYMMETRIC KEY MANAGEMENT}},
    month       = {Dec.},
    year        = {2023},
    number      = {CCSDS 354.0-M-1},
    institution = {Consultative Committee for Space Data Systems (CCSDS)},
    address     = {Washington, D.C., USA},
    url         = {https://ccsds.org/Pubs/354x0m1.pdf},
    note        = {{Recommended Practice, Issue 1}}
}

@techreport{CCSDS355.0-B-2,
    type = {Standard},
    title       = {{CCSDS Recommendation for Space Data System Practices: SPACE DATA LINK  SECURITY PROTOCOL}},
    month       = {Jul.},
    year        = {2022},
    number      = {CCSDS 355.0-B-2},
    institution = {Consultative Committee for Space Data Systems (CCSDS)},
    address     = {Washington, D.C., USA},
    url         = {https://ccsds.org/Pubs/355x0b2.pdf},
    note        = {{Recommended Standard, Issue 2}}
}

@techreport{CCSDS356.0-B-1,
    type = {Standard},
    title       = {{CCSDS Recommendation for Space Data System Practices: NETWORK LAYER  SECURITY  ADAPTATION PROFILE}},
    month       = {Jun.},
    year        = {2018},
    number      = {CCSDS 356.0-B-1},
    institution = {Consultative Committee for Space Data Systems (CCSDS)},
    address     = {Washington, D.C., USA},
    url         = {https://ccsds.org/Pubs/356xb1.pdf},
    note        = {{Recommended Standard, Issue 1}}
}

@techreport{CCSDS357.0-B-1,
    type = {Standard},
    title       = {{CCSDS Recommendation for Space Data System Practices: CCSDS  AUTHENTICATION  CREDENTIALS}},
    month       = {Jul.},
    year        = {2019},
    number      = {CCSDS 357.0-B-1},
    institution = {Consultative Committee for Space Data Systems (CCSDS)},
    address     = {Washington, D.C., USA},
    url         = {https://ccsds.org/Pubs/357x0b1.pdf},
    note        = {{Recommended Standard, Issue 1}}
}

@techreport{CCSDS357.1-O-1,
    type = {Standard},
    title       = {{Research and Development for
Space Data System Standards: INTERGOVERNMENTAL CERTIFICATION AUTHORITY}},
    month       = {Dec.},
    year        = {2024},
    number      = {CCSDS 357.1-O-1},
    institution = {Consultative Committee for Space Data Systems (CCSDS)},
    address     = {Washington, D.C., USA},
    url         = {https://ccsds.org/Pubs/357x1o1.pdf},
    note        = {{Experimental Specification, Issue 1}}
}

@misc{rfc3647,
    series =    {Request for Comments},
    number =    3647,
    howpublished =  {RFC 3647},
    publisher = {RFC Editor},
    doi =       {10.17487/RFC3647},
    url =       {https://www.rfc-editor.org/info/rfc3647},
    author =    {Warwick S. Ford and Santosh Chokhani and Stephen S. Wu and Randy V. Sabett and Charles (Chas) R. Merrill},
    title =     {{Internet X.509 Public Key Infrastructure Certificate Policy and Certification Practices Framework}},
    pagetotal = 94,
    year =      2003,
    month =     Nov,
}

@misc{rfc4158,
    series =    {Request for Comments},
    number =    4158,
    howpublished =  {RFC 4158},
    publisher = {RFC Editor},
    doi =       {10.17487/RFC4158},
    url =       {https://www.rfc-editor.org/info/rfc4158},
    author =    {Peter Hesse and Matt Cooper and Yuriy A. Dzambasow and Susan Joseph and Richard Nicholas},
    title =     {{Internet X.509 Public Key Infrastructure: Certification Path Building}},
    pagetotal = 81,
    year =      2005,
    month =     sep,
}

@misc{rfc6066,
    series =    {Request for Comments},
    number =     6066,
    howpublished =  {RFC 6066},
    publisher = {RFC Editor},
    doi =       {10.17487/RFC6066},
    url =       {https://www.rfc-editor.org/info/rfc6066},
    author =    {Donald E. Eastlake 3rd},
    title =     {{Transport Layer Security (TLS) Extensions: Extension Definitions}},
    pagetotal = 25,
    year =      2011,
    month =     jan,
}

@misc{rfc5055,
    series =    {Request for Comments},
    number =    5055,
    howpublished =  {RFC 5055},
    publisher = {RFC Editor},
    doi =       {10.17487/RFC5055},
    url =       {https://www.rfc-editor.org/info/rfc5055},
    author =    {Tim Polk and David Cooper and Russ Housley and Ambarish N. Malpani and Trevor Freeman},
    title =     {{Server-Based Certificate Validation Protocol (SCVP)}},
    pagetotal = 88,
    year =      2007,
    month =     dec,
    
}

@misc{rfc5280,
    series =    {Request for Comments},
    number =    5280,
    howpublished =  {RFC 5280},
    publisher = {RFC Editor},
    doi =       {10.17487/RFC5280},
    url =       {https://www.rfc-editor.org/info/rfc5280},
    author =    {Sharon Boeyen and Stefan Santesson and Tim Polk and Russ Housley and Stephen Farrell and David Cooper},
    title =     {{Internet X.509 Public Key Infrastructure Certificate and Certificate Revocation List (CRL) Profile}},
    pagetotal = 151,
    year =      2008,
    month =     may,
}

@misc{rfc3379,
    series =    {Request for Comments},
    number =    3379,
    howpublished =  {RFC 3379},
    publisher = {RFC Editor},
    doi =       {10.17487/RFC3379},
    url =       {https://www.rfc-editor.org/info/rfc3379},
    author =    {Russ Housley and Denis Pinkas},
    title =     {{Delegated Path Validation and Delegated Path Discovery Protocol Requirements}},
    pagetotal = 15,
    year =      2002,
    month =     sep,
}

@misc{rfc6960,
    series =    {Request for Comments},
    number =    6960,
    howpublished =  {RFC 6960},
    publisher = {RFC Editor},
    doi =       {10.17487/RFC6960},
    url =       {https://www.rfc-editor.org/info/rfc6960},
    author =    {Stefan Santesson and Michael Myers and Rich Ankney and Ambarish Malpani and Slava Galperin and Dr. Carlisle Adams},
    title =     {{X.509 Internet Public Key Infrastructure Online Certificate Status Protocol - OCSP}},
    pagetotal = 41,
    year =      2013,
    month =     jun,
}

@techreport{WEBurr1998,
   author = {W. E. Burr},
   title = {{Public Key Infrastructure (PKI) Technical Specifications: Part A Technical Concept of Operations}},
   institution = {National Institute of Standards and Technology},
   address= {},
   number = {NIST Working Draft TWG-98-59},
   DOI = {},
   url = {https://csrc.nist.rip/archive/pki-twg/baseline/pkicon20b.PDF},
   year = {1998},
}

@inproceedings{Jon2006,
  author    = {Jon Ølnes},
  title     = {{PKI Interoperability by an Independent, Trusted Validation Authority}},
  year      = {2006},
  booktitle = {5th Annual PKI R$\&$D Workshop “Making PKI Easy to Use”},
  url = {https://nvlpubs.nist.gov/nistpubs/Legacy/IR/nistir7313.pdf},
}

@techreport{ESAPSS-04-151,
    type = {Standard},
    title       = {{Telecommand Decoder Specification}},
    month       = {Sep.},
    year        = {1992},
    number      = {ESA PSS-04-151},
    institution = {European Space Agency (ESA)},
    address     = {Netherland},
    url         = {http://microelectronics.esa.int/vhdl/pss/PSS-04-151.pdf},
    note        = {{Issue 1}}
}

@techreport{ESAPSS-04-107,
    type = {Standard},
    title       = {{Packet Telecommand Standard}},
    month       = {Apr.},
    year        = {1992},
    number      = {ESA PSS-04-107},
    institution = {European Space Agency (ESA)},
    address     = {Netherland},
    url         = {http://microelectronics.esa.int/vhdl/pss/PSS-04-107.pdf},
    note        = {{Issue 2}}
}

@online{TDRS,
  title        = {{Tracking and Data Relay Satellites}},
  author       = {{National Aeronautics and Space Administration (NASA)}},
  month        = { },
  year         = { },
  url          = {https://www.nasa.gov/mission/tracking-and-data-relay-satellites/},
  urldate      = {August 15, 2024}
}

@online{TDRS-2023,
  title        = {{Tracking and Data Relay Satellite System Reimbursable for Fiscal Year 2024}},
  author       = {{National Aeronautics and Space Administration (NASA)}},
  month        = {Dec.},
  year         = {2023},
  url          = {https://www.nasa.gov/wp-content/uploads/2023/12/tdrs-reimbursable-rates-fy24-signed.pdf?emrc=434f1e},
  urldate      = {October 27, 2024}
}

@online{TDRS-capacity,
  title        = {{TDRS: TRACKING AND DATA RELAY SATELLITE
CONTINUING THE CRITICAL LIFELINE}},
  author       = {{National Aeronautics and Space Administration (NASA)}},
  month        = { },
  year         = { },
  url          = {https://www.nasa.gov/wp-content/uploads/2022/04/tdrsfactsheet_3.pdf?emrc=e97a55},
  urldate      = {December 05, 2024}
}

@online{SES-2025,
  title        = {{SES’s Ninth and Tenth O3b mPOWER Satellites Successfully Launched}},
  author       = {{Société Européenne des Satellites (SES)}},
  month        = {July},
  year         = {2025},
  url          = {https://www.ses.com/press-release/sess-ninth-and-tenth-o3b-mpower-satellites-successfully-launched},
  urldate      = {Jan 21, 2026}
}

@online{GPS-2024,
  title        = {{Global Positioning System (GPS) Overview}},
  author       = {{U.S. Department of Defense (DOD)}},
  month        = {Oct.},
  year         = {2024},
  url          = {https://www.navcen.uscg.gov/global-positioning-system-overview},
  urldate      = {October 22, 2024}
}

@online{ESA-2024,
  title        = {{Knowledge beyond our planet: space-based data centres}},
  author       = {{European Space Agency (ESA)}},
  month        = {Jul.},
  year         = {2024},
  url          = {https://www.esa.int/Enabling_Support/Preparing_for_the_Future/Discovery_and_Preparation/Knowledge_beyond_our_planet_space-based_data_centres?fbclid=IwY2xjawEgHgFleHRuA2FlbQIxMQABHb8yqgOrul9hQ7Og9TwxPHcoBN_BMCqpa11k-3rIlLHFHE7qrJ9xRilm8g_aem_Ux72gg-canpWHI2ddm6c7Q},
  urldate      = {July 8, 2024}
}

@online{ASCEND-2024,
  title        = {{Advanced Space Cloud for European Net zero Emission and Data sovereignty (ASCEND)}},
  author       = {{Thales Alenia Space}},
  month        = {},
  year         = {2024},
  url          = {https://ascend-horizon.eu/},
  urldate      = {November 11, 2024}
}

@online{Loft-2024,
  title        = {{Virtual Missions: Deploy your software on our space infrastructure}},
  author       = {{Loft Orbital}},
  month        = {},
  year         = {2024},
  url          = {https://www.loftorbital.com/fly-with-us/virtual-missions/},
  urldate      = {November 11, 2024}
}

@online{NASAOrbits-2009,
  title        = {{Catalog of Earth Satellite Orbits}},
  author       = {{National Aeronautics and Space Administration (NASA}},
  month        = {Sep.},
  year         = {2009},
  url          = {https://earthobservatory.nasa.gov/features/OrbitsCatalog},
  urldate      = {October 29, 2024}
}

@techreport{BCA,
  title        = {{Bridge Certification Authorities: Connecting B2B Public Key Infrastructures}},
  author       = {William T. Polk and Nelson E. Hastings},
  month        = {},
  year         = {},
  institution = {{National Institute of Standards and Technology}},
  url          = {https://csrc.nist.rip/groups/ST/crypto_apps_infra/documents/B2B-article.pdf},
  urldate      = {December 02, 2024}
}

@article{ALTERMAN2001,
title = {{The US Federal PKI and the Federal Bridge Certification Authority}},
journal = {Computer Networks},
volume = {37},
number = {6},
pages = {685-690},
year = {2001},
issn = {1389-1286},
doi = {https://doi.org/10.1016/S1389-1286(01)00244-4},
url = {https://www.sciencedirect.com/science/article/pii/S1389128601002444},
author = {Peter Alterman}
}

@online{CCSDS,
  title        = {{Consultative Committee for Space Data Systems (CCSDS)}},
  url          = {https://ccsds.org/},
  urldate      = {January 29, 2026}
}

@article{VONMAURICH2018,
title = {Data authentication, integrity and confidentiality mechanisms for federated satellite systems},
journal = {Acta Astronautica},
volume = {149},
pages = {61-76},
year = {2018},
issn = {0094-5765},
doi = {https://doi.org/10.1016/j.actaastro.2018.05.003},
url = {https://www.sciencedirect.com/science/article/pii/S0094576517301418},
author = {Olga {von Maurich} and Alessandro Golkar},
}

@article{Neish2019,
author = {Neish, Andrew and Walter, Todd and Powell, J. David},
title = {Design and analysis of a public key infrastructure for SBAS data authentication},
journal = {NAVIGATION},
volume = {66},
number = {4},
pages = {831-844},
doi = {https://doi.org/10.1002/navi.338},
year = {2019}
}

@article{Yates2023,
author = {John Yates},
title = {{The impact of weather on Ka-band frequencies}},
journal = {{ROOM - Space Journal of Asgardia}},
number = {33},
url = {https://room.eu.com/article/the-impact-of-weather-on-ka-band-frequencies},
year = {2023}
}

@MISC{ESA,
author = {{ESA - The European Space Agency}},
title  = {Satellite frequency bands},
url    = {https://www.esa.int/Applications/Connectivity_and_Secure_Communications/Satellite_frequency_bands},
note   = {[Accessed: 01-Feb-2026]},
 }

@BOOK{Book-2008,
  title={Satellite Communications Systems Engineering: Atmospheric Effects, Satellite Link Design, and System Performance},
  author = {Louis J. Ippolito},
  edition={1st},
  year = {2008}, 
  publisher = {Wiley},
}

@misc{keyspace,
      title={{KeySpace: Enhancing Public Key Infrastructure for Interplanetary Networks}}, 
      author={Joshua Smailes and Filip Futera and Sebastian Köhler and Simon Birnbach and Martin Strohmeier and Ivan Martinovic},
      year={2026},
      eprint={2408.10963},
      archivePrefix={arXiv},
      primaryClass={cs.CR},
      url={https://arxiv.org/abs/2408.10963}, 
}

@INPROCEEDINGS{ADOPT_OCSP,
  author={Marias, G. F. and Papapanagiotou, K. and Georgiadis, P.},
  booktitle={11th European Wireless Conference 2005 - Next Generation wireless and Mobile Communications and Services}, 
  title={{ADOPT. A Distributed OCSP for Trust Establishment in MANETs}}, 
  year={2005},
  volume={},
  number={},
  pages={1-7},
}

@inproceedings{delta_CRLs,
  author    = {David A. Cooper},
  title     = {A More Efficient Use of Delta-CRLs},
  booktitle = {Proceedings of the 2000 IEEE Symposium on Security and Privacy},
  pages     = {190--202},
  year      = {2000},
  publisher = {IEEE},
  url       = {https://nist.gov}
}

@article{Merkle_Tree,
author = {Mu\~{n}oz, Jose L. and Forne, Jordi and Esparza, Oscar and Soriano, Miguel},
title = {{Certificate revocation system implementation based on the Merkle hash tree}},
year = {2004},
issue_date = {Jan. 2004},
publisher = {Springer-Verlag},
address = {Berlin, Heidelberg},
volume = {2},
number = {2},
issn = {1615-5262},
journal = {Int. J. Inf. Secur.},
month = jan,
pages = {110–124},
numpages = {15},
}

@inproceedings{DTN,
author = {Fall, Kevin},
title = {{A Delay-Tolerant Network Architecture for Challenged Internets}},
year = {2003},
isbn = {1581137354},
publisher = {Association for Computing Machinery},
address = {New York, NY, USA},
url = {https://doi.org/10.1145/863955.863960},
doi = {10.1145/863955.863960},
booktitle = {Proceedings of the 2003 Conference on Applications, Technologies, Architectures, and Protocols for Computer Communications},
pages = {27–34},
numpages = {8},
location = {Karlsruhe, Germany},
series = {SIGCOMM '03}
}

@INPROCEEDINGS{CRLite,
  author={Larisch, James and Choffnes, David and Levin, Dave and Maggs, Bruce M. and Mislove, Alan and Wilson, Christo},
  booktitle={2017 IEEE Symposium on Security and Privacy (SP)}, 
  title={{CRLite: A Scalable System for Pushing All TLS Revocations to All Browsers}}, 
  year={2017},
  volume={},
  number={},
  pages={539-556},
  doi={10.1109/SP.2017.17}}

@INPROCEEDINGS{Revoc,
  author={Koisser, David and Schwarzkopf, Albert and Brasser, Ferdinand and Da Broi, Giacomo},
  booktitle={2025 Security for Space Systems (3S)}, 
  title={{Efficient PKI Design for Secure Communication and Collaboration in Space Networks}}, 
  year={2025},
  volume={},
  number={},
  pages={1-12},
}

@phdthesis{DHT,
  author = {Frank Dabek},
  title  = {{A Distributed Hash Table}},
  school = {Massachusetts Institute of Technology},
  year   = {2005},
  url    = {https://pdos.csail.mit.edu/papers/fdabek-phd-thesis.pdf}
}

@inproceedings{CCN,
author = {Jacobson, Van and Smetters, Diana K. and Thornton, James D. and Plass, Michael F. and Briggs, Nicholas H. and Braynard, Rebecca L.},
title = {{Networking Named Content}},
year = {2009},
isbn = {9781605586366},
publisher = {Association for Computing Machinery},
address = {New York, NY, USA},
doi = {10.1145/1658939.1658941},
booktitle = {Proceedings of the 5th International Conference on Emerging Networking Experiments and Technologies},
pages = {1–12},
numpages = {12},
location = {Rome, Italy},
series = {CoNEXT '09}
}

@article{NDN,
author = {Zhang, Lixia and Afanasyev, Alexander and Burke, Jeffrey and Jacobson, Van and claffy, kc and Crowley, Patrick and Papadopoulos, Christos and Wang, Lan and Zhang, Beichuan},
title = {Named Data Networking},
year = {2014},
issue_date = {July 2014},
publisher = {Association for Computing Machinery},
address = {New York, NY, USA},
volume = {44},
number = {3},
issn = {0146-4833},
doi = {10.1145/2656877.2656887},
journal = {SIGCOMM Comput. Commun. Rev.},
month = jul,
pages = {66–73},
numpages = {8},
}

@techreport{NDNCert,
  author      = {Zhiyi Zhang and Yingdi Yu and Alexander Afanasyev and Lixia Zhang},
  title       = {{NDN Certificate Management Protocol (NDNCERT)}},
  institution = {Named Data Networking (NDN)},
  number      = {NDN-0050},
  year        = {2017},
  url         = {https://named-data.net/publications/techreports/ndn-0050-1-ndncert/}
}

@inproceedings{DistributedOCSP,
author = {Rezende, Daniel and Maziero, Carlos and Mannes, Elisa},
title = {{A Distributed Online Certificate Status Protocol for Named Data Networks}},
year = {2018},
isbn = {9781450351911},
publisher = {Association for Computing Machinery},
address = {New York, NY, USA},
doi = {10.1145/3167132.3167358},
booktitle = {Proceedings of the 33rd Annual ACM Symposium on Applied Computing},
pages = {2102–2108},
numpages = {7},
location = {Pau, France},
series = {SAC '18}
}

@InProceedings{ThresholdDSS,
author={Gennaro, Rosario and Jarecki, Stanis{\l}aw and Krawczyk, Hugo
and Rabin, Tal},
editor={Maurer, Ueli},
title={Robust Threshold DSS Signatures},
booktitle={Advances in Cryptology --- EUROCRYPT '96},
year={1996},
publisher={Springer Berlin Heidelberg},
address={Berlin, Heidelberg},
pages={354--371},
}

@article{COCA,
author = {Zhou, Lidong and Schneider, Fred B. and Van Renesse, Robbert},
title = {{COCA: A secure distributed online certification authority}},
year = {2002},
issue_date = {November 2002},
publisher = {Association for Computing Machinery},
address = {New York, NY, USA},
volume = {20},
number = {4},
issn = {0734-2071},
doi = {10.1145/571637.571638},
journal = {ACM Trans. Comput. Syst.},
month = nov,
pages = {329–368},
numpages = {40},
}

@INPROCEEDINGS{QM_Proto,
  author={Zefreh, Mohammad Sheikh and Fanian, Ali and Sajadieh, Sayyed Mahdi and Berenjkoub, Mahdi and Khadivi, Pejman},
  booktitle={2008 10th International Conference on Advanced Communication Technology}, 
  title={{A Distributed Certificate Authority and Key Establishment Protocol for Mobile Ad Hoc Networks}}, 
  year={2008},
  volume={2},
  number={},
  pages={1157-1162},
  doi={10.1109/ICACT.2008.4493971}
  }

@inproceedings{PBFT,
author = {Castro, Miguel and Liskov, Barbara},
title = {{Practical Byzantine Fault Tolerance}},
year = {1999},
isbn = {1880446391},
publisher = {USENIX Association},
address = {USA},
booktitle = {Proceedings of the Third Symposium on Operating Systems Design and Implementation},
pages = {173–186},
numpages = {14},
location = {New Orleans, Louisiana, USA},
series = {OSDI '99}
}

@inproceedings{PBFT2,
author = {Yin, Maofan and Malkhi, Dahlia and Reiter, Michael K. and Gueta, Guy Golan and Abraham, Ittai},
title = {{HotStuff: BFT Consensus with Linearity and Responsiveness}},
year = {2019},
isbn = {9781450362177},
publisher = {Association for Computing Machinery},
address = {New York, NY, USA},
doi = {10.1145/3293611.3331591},
booktitle = {Proceedings of the 2019 ACM Symposium on Principles of Distributed Computing},
pages = {347–356},
numpages = {10},
location = {Toronto ON, Canada},
series = {PODC '19}
}

@inproceedings{BFT,
author = {Miller, Andrew and Xia, Yu and Croman, Kyle and Shi, Elaine and Song, Dawn},
title = {{The Honey Badger of BFT Protocols}},
year = {2016},
isbn = {9781450341394},
publisher = {Association for Computing Machinery},
address = {New York, NY, USA},
doi = {10.1145/2976749.2978399},
booktitle = {Proceedings of the 2016 ACM SIGSAC Conference on Computer and Communications Security},
pages = {31–42},
numpages = {12},
location = {Vienna, Austria},
series = {CCS '16}
}

@misc{rfc5934,
    series =    {Request for Comments},
    number =    5934,
    howpublished =  {RFC 5934},
    publisher = {RFC Editor},
    doi =       {10.17487/RFC5934},
    url =       {https://www.rfc-editor.org/info/rfc5934},
    author =    {Carl Wallace and Sam Ashmore and Russ Housley},
    title =     {{Trust Anchor Management Protocol (TAMP)}},
    pagetotal = 91,
    year =      2010,
    month =     aug,
}

@article{CRYSTALS-Dilithium, 
title={{CRYSTALS-Dilithium: A Lattice-Based Digital Signature Scheme}}, volume={2018}, 
url={https://tches.iacr.org/index.php/TCHES/article/view/839}, 
DOI={10.13154/tches.v2018.i1.238-268}, 
number={1}, 
journal={IACR Transactions on Cryptographic Hardware and Embedded Systems}, 
author={Ducas, Léo and Kiltz, Eike and Lepoint, Tancrède and Lyubashevsky, Vadim and Schwabe, Peter and Seiler, Gregor and Stehlé, Damien}, 
year={2018}, 
month={Feb.}, 
pages={238–268} 
}

@techreport{CBOR,
    number =    {draft-ietf-cose-cbor-encoded-cert-17},
    type =      {Internet-Draft},
    institution =   {Internet Engineering Task Force},
    publisher = {Internet Engineering Task Force},
    note =      {Work in Progress},
    url =       {https://datatracker.ietf.org/doc/draft-ietf-cose-cbor-encoded-cert/17/},
    author =    {John Preuß Mattsson and Göran Selander and Shahid Raza and Joel Höglund and Martin Furuhed},
    title =     {{CBOR Encoded X.509 Certificates (C509 Certificates)}},
    pagetotal = 88,
    year =      2026,
    month =     mar,
    day =       2,
}

\appendix

\section{Public Key Infrastructure} 
\label{apndx:PKI}

A Public Key Infrastructure (PKI) provides a secure framework for managing digital certificates and public-private key pairs in public key cryptographic systems. The main components of PKI include \cite{rfc3647,rfc4158,rfc6960}:

\begin{itemize}

    \item \textit{Registration Authority (RA)}: Verifies the identity (name, organization, domain, etc.) of users or devices requesting digital certificates, either online or offline. This role is optional and can be performed by the CA or by a third party.
    
    \item \textit{Certificate Authority (CA)}: A trusted entity that vouches for the binding between the end-user's identity (name, organization, domain, etc.) and its data items (e.g., public key) in a certificate by digitally signing the certificate. The CA takes care of the following responsibilities:
     
    \begin{enumerate}
        \item \textit{Certificate Issuance.} A CA, after identity verification of the end certificate user, issues a signed certificate to the user with a set validity period.
     
        \item \textit{Certificate Renewal.} When a certificate is nearing expiration, the CA may issue a new certificate (renewal) based on the request from the end certificate user.
 
        \item \textit{Certificate Revocation.} A CA may revoke a certificate in some cases, for instance, if certificate or user's private key is compromised, user's status has changed, or certificate is no longer needed. CA provides the up-to-date revocation status of the certificates using the Online Certificate Status Protocol (OCSP), Certificate Revocation Lists (CRLs), or some other mechanism.
    \end{enumerate}    

    \item \textit{Certificate Repository (CR)}: A secure system or collection of distributed systems that stores and manages certificates and CRLs, and facilitates the distribution of these certificates and CRLs to end entities. CAs post certificates and CRLs to repositories.

    \item \textit{Certificate Policy (CP)}: A policy that defines the procedures of a PKI and offers guidelines for external parties to assess its trustworthiness. A CP is a set of rules specifying the applicability of a certificate to a particular group or application with common security requirements.

    \item \textit{Certificate Validation Process (CVP)}: A process that verifies that the certificate is valid, active and has not expired, that the certificate has integrity and has not been altered or tampered with and that the certificate has not been revoked by the CA.

    \item \textit{Relying Party (RP)}: A certificate user that needs to validate the received certificate before relying on the authentication, non-repudiation, or confidentiality services associated with the public key in that particular certificate. The RP takes care of the following tasks:
    \begin{enumerate}
        \item \textit{Certificate Status Validation.} The RP first verifies the digital signature of the certificate to ensure it was issued by a trusted CA. It then checks the certificate's current status to determine if it has been revoked or expired using CRLs or OCSP. To reduce lookup overhead, RPs may cache certificates and CRLs.
     
        \item \textit{Compliance Checking.} The RP further checks the compliance of certificate against the policies established by the CA, including the key use restrictions.
        
        \item \textit{Certificate Path Discovery \& Validation.} In case of hierarchical PKI architecture, the RP builds the complete certificate path when it’s not directly available and performs certificate path validation by validating the entire certificate chain, including checking revocation statuses and policy compliance.    
    \end{enumerate}

\end{itemize}

\subsection{PKI Architectures} 

A PKI architecture defines how trust relationships are organized and managed within a network, tailored to the system’s complexity, scalability, and interoperability requirements.

\begin{figure}[h]
    \centering
    \includegraphics[width=1\textwidth]{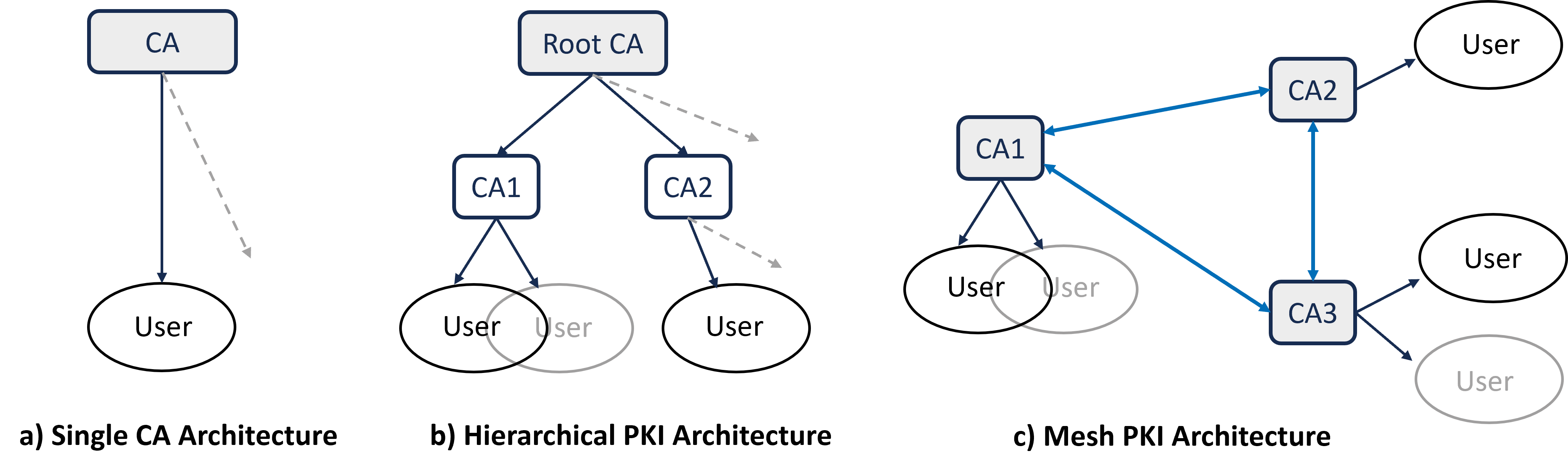}
    \caption{Different PKI Architectures}
    \label{fig:pki}
\end{figure}

\begin{itemize}
    \item \textbf{Single CA:} A basic PKI setup where one CA issues, renews, and revokes certificates for all users, serving as the sole trust anchor (Figure~\ref{fig:pki}a).
     
    \item \textbf{Hierarchical PKI:} In this architecture, trust flows from a root CA to intermediate CAs and then to end users, with each entity trusting the one above it (Figure~\ref{fig:pki}b).
     
    \item \textbf{Mesh PKI:} A peer-to-peer PKI where CAs cross-certify each other, forming a “web of trust” based on predefined policies (Figure~\ref{fig:pki}c).
\end{itemize}

\end{document}